\documentclass[english,aps,prd,nofootinbib,twocolumn]{revtex4-1}
\usepackage{CJK}

\usepackage{amsfonts}
\usepackage{amsmath}
\usepackage{hyperref}
\usepackage{amssymb}
\usepackage{cleveref}
\usepackage{cancel}
\usepackage{xfrac}
\usepackage{xcolor}
\usepackage{relsize}
\usepackage[bottom]{footmisc}

\usepackage{tikz}
\usepackage{tikz-feynman}
\usetikzlibrary{calc,arrows,decorations.markings,snakes,shapes}
\usepackage{pgfplots}
\pgfplotsset{compat=1.3}
\usetikzlibrary{3d,calc}

\usepackage{units}

\usepackage{xspace}

\usepackage{color}
\usepackage{graphicx}
\usepackage[space]{grffile}
\usepackage{verbatim}
\usepackage{amsmath}
\usepackage{amssymb}
\usepackage{wasysym}
\usepackage[caption=false]{subfig}
\usepackage{url}
\usepackage{bbold}
\usepackage{slashed}
\usepackage{epstopdf}
\usepackage{braket}
\usepackage{float}

\DeclareMathAlphabet\mathbfcal{OMS}{cmsy}{b}{n}

\newcommand{\boundellipse}[3]
{[black,fill=blue!30] (#1) ellipse (#2 and #3)
}
\newcommand{\boundellipseW}[3]
{[white,fill=white] (#1) ellipse (#2 and #3)
}
\newcommand*{\Scale}[2][4]{\scalebox{#1}{$#2$}}%

\begin{document}

\title{Bosonization of the $\mathbf{Q}=0$ continuum of Dirac Fermions}

\author{Sebastian Mantilla$^{1}$}
\author{Inti Sodemann$^{1}$}


\affiliation{$^1$Max-Planck Institute for the Physics of Complex Systems, D-01187 Dresden, Germany}

\date{\today}



\begin{abstract}
We develop a bosonization formalism that captures non-perturbatively the interaction effects on the $\mathbf{Q}=0$ continuum of excitations of nodal fermions above one dimension. Our approach is a natural extension of the classic bosonization scheme for higher dimensional Fermi surfaces 
to include the $\mathbf{Q}=0$ neutral excitations that would be absent in a single-band system. The problem is reduced to solving a boson bilinear Hamiltonian. We establish a rigorous microscopic footing for this approach by showing that the solution of such boson bilinear Hamiltonian is exactly equivalent to performing the infinite sum of Feynman diagrams associated with the Kadanoff-Baym particle-hole propagator that arises from the self-consistent Hartree-Fock approximation to the single particle Green's function. We apply this machinery to compute the interaction corrections to the optical conductivity of 2D Dirac Fermions with Coulomb interactions reproducing the results of perturbative renormalization group at weak coupling and extending them to the strong coupling regime. 
\end{abstract}

\maketitle


\textit{\color{blue} Introduction}. The remarkable success of bosonization in capturing the non-perturbative properties of interacting fermions in one-dimension \cite{giamarchi2003quantum} has long motivated the quest for extensions of this program to higher dimensions. One major such enterprise has been the development of higher dimensional bosonization of Fermi surfaces \cite{luther1979tomonaga, haldane2005luttinger, houghton1993bosonization, neto1994bosonization,houghton2000multidimensional}. In this approach, particle-hole creation operators of a given total momentum $\mathbf{Q}$, $c^{\textcolor{black}{\dagger}}_{\mathbf{k}+\mathbf{Q}/2}c^{\textcolor{white}{\dagger}}_{\mathbf{k}-\mathbf{Q}/2}$, are promoted to bosonic creation operators with a commutator that is approximated as a number. The resulting bosonized Hamiltonian only couples bosonic modes with momentum $\mathbf{Q}$ to bosons with either $+\mathbf{Q}$ or $-\mathbf{Q}$. Namely, there is zero amplitude for a particle hole-pair with momentum $\mathbf{Q}$ to transition into two particle-hole pairs with momenta $\mathbf{Q}_{1,2}$ and $\mathbf{Q}=\mathbf{Q}_1+\mathbf{Q}_2$. The only allowed process are for the particle-hole pair with momentum $\mathbf{Q}$ to scatter into another one with the same $\mathbf{Q}$, or to create pairs of particle-hole pairs with momentum $+\mathbf{Q}$ and $-\mathbf{Q}$ (for a succinct incarnation of this structure see, e.g., Eq.(7.1) in Ref. \cite{neto1995exact}). This assumption of separability of Hilbert spaces of particle hole pairs with different magnitudes of $|\mathbf{Q}|$, lies at the heart of the higher dimensional bosonization approach to Fermi surfaces and it is believed to be an asymptotically correct description of particle-hole excitations of Landau fermi liquids at small $|\mathbf{Q}|$.

\begin{figure}
\centering	
\includegraphics[scale=0.48,page=1]{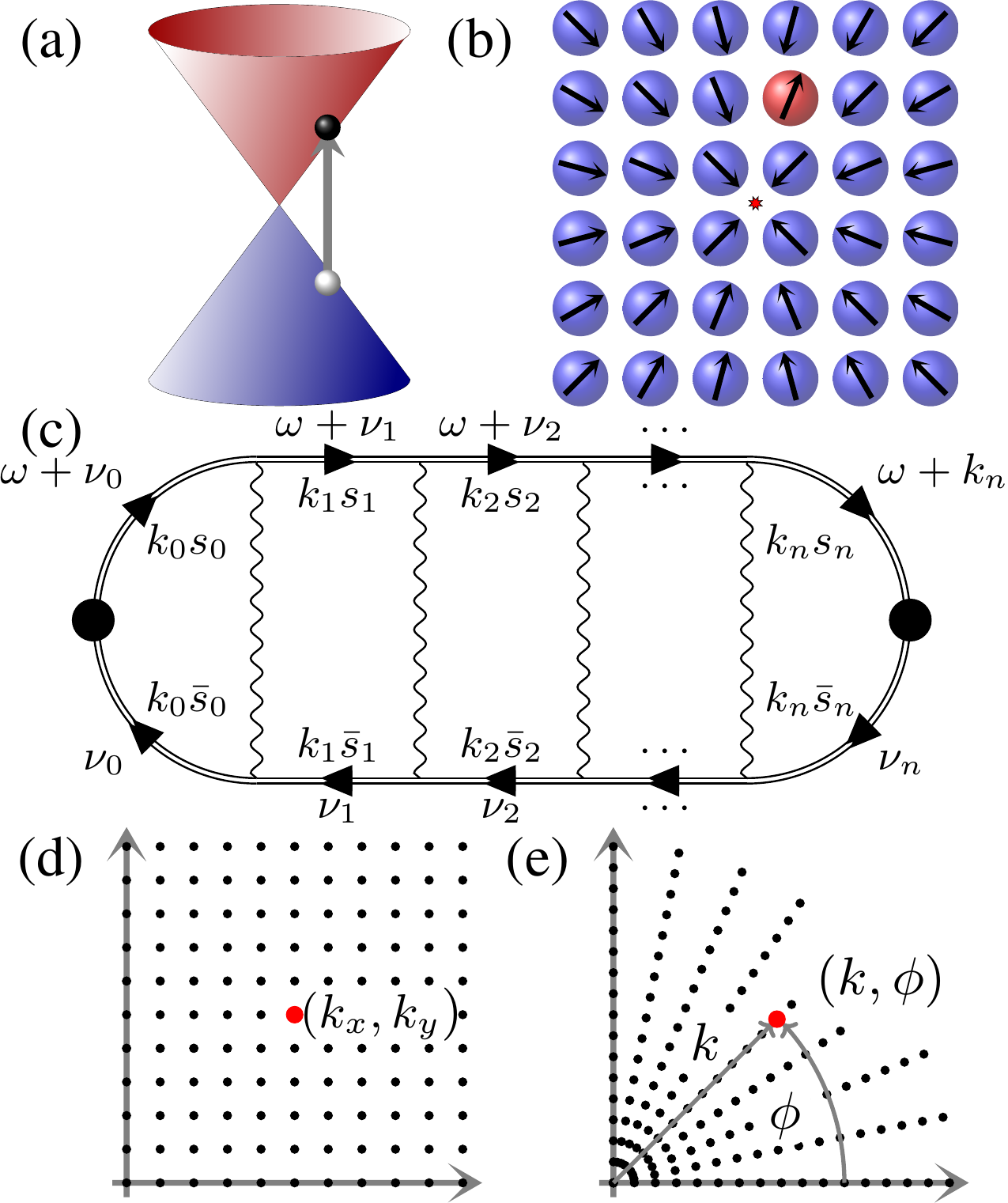}
\caption{(a,b) Creation of an electron-hole pair as the flipping of a pseudospin on a vortex configuration. (c) The $n$-th order diagram of the KB sum for the particle-hole propagator associated with SCHF. The double lines are Green's functions dressed by SCHF self-energies, and the wiggly lines are interaction matrices. (d,e) Rediscretization from the square to the polar lattice.}
\label{fig:Dibujos}
\end{figure}

Ordinary single-band Fermi liquids do not have low energy particle-hole excitations with total momentum $\mathbf{Q}=0$ and therefore this sector does not appear in the conventional problem of bosonization of Fermi surfaces. In contrast, nodal semimetals, in which the Fermi surface shrinks to a point, such as Weyl or massless Dirac semimetals, have a non-trivial set of gapless optical $\mathbf{Q}=0$ particle-hole excitations. The central purpose of the present study is to develop a systematic bosonization approach to this sector for gapless semimetals. For concreteness we will discuss only 2D massless Dirac fermions, such as those appearing in graphene and the surface of 3D topological insulators, but our ideas can be naturally extended to other cases and higher dimensions. To describe such excitations, we will borrow the central assumption of the bosonization approach of Fermi surfaces, namely, that such optical particle-hole pairs are decoupled from the particle-hole pairs of finite momentum $\mathbf{Q}$. We expect this simplification to be justified at low energies in phases which are adiabatically related to free fermions, in a similar sense to how such decoupling allows to describe Fermi liquids which are adiabatically related to free fermions in the higher dimensional bosonization of Fermi surfaces. We will, however, establish, a very explicit and solid connection between our bosonization approach and the conventional Feynman diagrammatic perturbation theory that demonstrates the validity of this central assumption of our approach. Specifically, we will prove that the solution of our effective bosonic Hamiltonian for the optical particle-hole pairs is \textit{exactly equivalent} to the self-consistent Kadanoff-Baym resummation~\cite{baym1961conservation, PhysRev.127.1391} of the particle-hole propagator at $\mathbf{Q}=0$, associated with the self-consistent Hartree-Fock approximation to the single particle-particle Green's function. 

As an application of our approach we will compute the interaction corrections to the optical conductivity of 2D Dirac fermions with Coulomb interactions, whose strength is parametrized by the effective fine structure constant $\alpha = e^2/ \epsilon v$, where $v$ is the velocity of the Dirac fermions and $\epsilon$ the dielectric constant of the surrounding medium. This optical conductivity at low energies is determined by fundamental constants of nature, and given by $\sigma_0=e^2/16 \hbar$ per Dirac cone~\cite{PhysRevB.50.7526, ando2002dynamical}. Its zero frequency limit is not expected to be renormalized by interactions, but, Coulomb interactions can produce a slow flow as a function of frequency to such value and a non-trivial non-analytic frequency dependence at low energies. Early perturbative calculations of such corrections where in mutual disagreement~\cite{mishchenko2008minimal, herbut2008coulomb}, but subsequent studies~\cite{sheehy2009optical, abedinpour2011drude, sodemann2012interaction, gazzola2013conductivity, barnes2014effective, teber2014interaction, teber2018field} validated the result of Ref.~\cite{mishchenko2008minimal}. As we will see, our approach will recover the perturbative results of Ref.~\cite{mishchenko2008minimal} at small interactions and extend them non-perturbatively to finite $\alpha$. Non-perturbative attempts to understand the effects of Coulomb interactions in the optical conductivity of Dirac fermions have been scarce. A Quantum Monte Carlo effort~\cite{boyda2016many} to compute the optical conductivity concluded that interaction corrections remain rather small even at $\alpha \sim 2$. Our analysis will also support this conclusion, which is broadly in agreement with experiments that have found values close to that for non-interacting fermions~\cite{li2008dirac, mak2008measurement, nair2008fine}.

\textit{\color{blue} Effective Hamiltonian and Hilbert space}. The microscopic Hamiltonian is ($\hbar = 1$):
\begin{equation}
\label{eq:Main-Ham}
\begin{split}
&H = 
v\sum_{\mathbf{k },\sigma,\sigma'}
\psi_{\mathbf{k },\sigma}^{\textcolor{black}{\dagger}}
\left( \mathbf{k} \cdot 
\boldsymbol{\sigma}_{\sigma\sigma'} \right)
\psi_{\mathbf{k },\sigma'}^{\textcolor{white}{\dagger}}\\
&+ \frac{1}{2A}
\sum_{\mathbf{k }\mathbf{k'}}
\sum_{\sigma \sigma'} 
V_{\mathbf{q}}
\psi_{\mathbf{k'+q,\sigma'}}^{\textcolor{black}{\dagger}}
\psi_{\mathbf{k -q,\sigma }}^{\textcolor{black}{\dagger}}
\psi_{\mathbf{k   ,\sigma }}^{\textcolor{white}{\dagger}}
\psi_{\mathbf{k'  ,\sigma'}}^{\textcolor{white}{\dagger}},
\end{split}
\end{equation}
where $A$ is the system area, and $V_{\mathbf{q}}$ is the Fourier transform of the interaction potential.  It is convenient to imagine the Fermions moving in a 2D Torus so that its momentum is quantized on a lattice. In this momentum lattice the complete many-body Hilbert space is a tensor product of empty, singly and doubly occupied states:
\begin{equation}
    \mathcal{H} = \bigotimes_{\mathbf{k}}
    \left (
    \left| 0          \right\rangle_{\mathbf{k}} \oplus 
    \left| \uparrow   \right\rangle_{\mathbf{k}} \oplus 
    \left| \downarrow \right\rangle_{\mathbf{k}} \oplus 
    \left| \uparrow  \downarrow \right\rangle_{\mathbf{k}}
    \right ).
\end{equation}
The kinetic term in Eq. \eqref{eq:Main-Ham} produces no fluctuations between the occupancy of the momentum sites, and favors a ground state with singly occupied states with a suitably oriented spin in the form of vortex around the Dirac point (see Fig. \ref{fig:Dibujos}a). The interactions are a form of pair hopping terms that have a finite amplitude to induce transitions into states with doubly occupied sites and empty sites. Crucially, the subspace of the Hilbert space with singly occupied sites is equivalent to the space of particle-hole pairs with zero total momentum, ${\bf Q}=0$, while those states with doubly occupied and empty sites contain particle-hole excitations of finite momentum ${\bf Q}$. Therefore, following the spirit of higher dimensional bosonization, we will project the Hamiltonian in Eq. \eqref{eq:Main-Ham} onto the Hilbert space of singly occupied sites in the momentum lattice, depicted in Fig. \ref{fig:Dibujos}b. This Hilbert space contains a spin-1/2 at each momentum site:
\begin{equation}
\label{eq:Hilbert-space-Spin}
    \mathcal{H}_{\mathrm{single}} = \bigotimes_{\mathbf{k}}
    \left (
    \left| \uparrow   \right\rangle_{\mathbf{k}} \oplus 
    \left| \downarrow \right\rangle_{\mathbf{k}} 
    \right ),
\end{equation}
and the projection of the Hamiltonian from Eq. \eqref{eq:Main-Ham} leads to the following Heisenberg model:
\begin{equation}
\label{eq:Effective-Heisenberg-Hamiltonian}
\begin{split}
\mathcal{P} H \mathcal{P} &= 
\sum_{\mathbf{k }} v \mathbf{k }\cdot \mathbf{s}_{\mathbf{k }} - 
\sum_{\mathbf{k} \neq \mathbf{k'}}
\frac{V_{\mathbf{k}-\mathbf{k'}}}{4A}
\mathbf{s}_{\mathbf{k }}
\cdot
\mathbf{s}_{\mathbf{k'}}
,
\end{split}
\end{equation}
where $\mathbf{s}_{\mathbf{k}}=\sum_{\sigma,\sigma'}
\psi_{\mathbf{k },\sigma}^{\textcolor{black}{\dagger}}
\boldsymbol{\sigma}_{\sigma\sigma'} 
\psi_{\mathbf{k },\sigma'}^{\textcolor{white}{\dagger}}$ is a spin operator for the $\mathbf{k}$ site of the momentum lattice. The first term in Eq. \eqref{eq:Effective-Heisenberg-Hamiltonian} is a Zeeman vortex field and the second term is a long-range exchange coupling. 

This Hamiltonian is not exactly solvable but the fluctuations around the non-interacting state can be described by a Holstein-Primakoff expansion \cite{auerbach2012interacting}. To do so, we choose a spin basis that diagonalizes the kinetic energy at each momentum site $
\mathbf{s}_{\mathbf{k}} = -
s_{\mathbf{k}}^{z}\hat{\mathbf{k}} + 
s_{\mathbf{k}}^{x} \hat{\mathbf{z}} + 
s_{\mathbf{k}}^{y}\hat{\boldsymbol{\phi}}$
where $\hat{\mathbf{z}}$ is the out-of-plane direction and $\hat{\boldsymbol{\phi}} = \hat{\mathbf{z}} \times \hat{\mathbf{k}}$. 
The spin operators can be expanded as 
$
s_{\mathbf{k }}^{z} \approx 1 - 2 
b_{\mathbf{k }}^{\textcolor{black}{\dagger}}
b_{\mathbf{k }}^{\textcolor{white}{\dagger}}$, $
s_{\mathbf{k }}^{x} \approx 
b_{\mathbf{k }}^{\textcolor{white}{\dagger}} + 
b_{\mathbf{k }}^{\textcolor{black}{\dagger}}$, and $i s_{\mathbf{k }}^{y} \approx 
b_{\mathbf{k }}^{\textcolor{white}{\dagger}} - 
b_{\mathbf{k }}^{\textcolor{black}{\dagger}}$. 
Up to boson bilinears the Hamiltonian becomes (see \S\ref{sect:Supp:Holstein-Primakoff} of \cite{SupplementalMaterial}):
\begin{equation}
\label{eq:HP-Hamiltonian}
\begin{split}
&H_{HP} = 
\sum_{\mathbf{k},\mathbf{k'}} 
B_{\mathbf{k }}^{\textcolor{black}{\dagger}}
H_{\mathbf{k }\mathbf{k'}}
B_{\mathbf{k'}}^{\textcolor{white}{\dagger}}, 
\end{split}
\end{equation}
with
$B_{\mathbf{k }}^{\textcolor{black}{\dagger}} =
\begin{pmatrix}
b_{\mathbf{k }}^{\textcolor{black}{\dagger}}
&&
b_{\mathbf{k }}^{\textcolor{white}{\dagger}}
\end{pmatrix}$, and 
\begin{equation}
\label{eq:HP-Explicit-Hamiltonian}
\begin{split}
H_{\mathbf{k k'}} &= 
\delta_{\mathbf{k k'}}
\begin{pmatrix}
2 E_{\mathbf{k}}	&	0	\\
0   &   -2 E_{\mathbf{k}}
\end{pmatrix} - T_{\mathbf{k k'}},
\end{split}
\end{equation}
with $E_{\mathbf{k}} = v|\mathbf{k}| + \Sigma_{\mathbf{k}}$ and $\Sigma_{\mathbf{k}} = \sum_{ \mathbf{k'}} 
V_{\mathbf{k-k'}}\cos \phi_{\mathbf{kk'}}/2A$ is the Hartree-Fock self-energy and $T_{\mathbf{kk'}}$ is the interaction matrix in the band basis (for details see \S\ref{sect:Supp:Band-basis} of \cite{SupplementalMaterial}) 
\begin{equation}
\label{eq:Interacion-matrix}
T_{\mathbf{kk'}} = 
\frac{V_{\mathbf{k-k'}}}{4A} 
\begin{pmatrix}
    1+\cos\phi_{\mathbf{kk'}}  &    1-\cos\phi_{\mathbf{kk'}} \\  
    1-\cos\phi_{\mathbf{kk'}}  &    1+\cos\phi_{\mathbf{kk'}}
\end{pmatrix} .   
\end{equation}

\textit{\color{blue} Connection to perturbation theory}. 
We will now demonstrate that the solution of the boson bilinear Hamiltonian in Eq. \eqref{eq:HP-Hamiltonian} is exactly equivalent to the calculation of the particle-hole propagator within the Kadanoff-Baym (KB) resummation of Feynman diagrams associated with the self-consistent Hartree-Fock (SCHF) approximation to the single-particle Green's function. In terms of electrons, the boson creation operator, $b^{\dagger}_{\mathbf{k}}$ corresponds to the interband $\mathbf{Q}=0$ electron-hole pair creation operator: $b^{\dagger}_{\mathbf{k}}=\psi^{\textcolor{black}{\dagger}}_{\mathbf{k}+}\psi^{\textcolor{white}{\dagger}}_{\mathbf{k}-}$, where the subindex $s=\pm$ denotes the conduction and valence bands. Therefore, our goal is to compute the electron-hole pair propagator defined as:
\begin{equation}
\label{eq:Particle-hole-propagator}
\begin{split}
\chi_{\mathbf{k}_{1}\mathbf{k}_{2}}^{s_{1}s_{2}}(t) &= -iT\left\langle
\psi_{\mathbf{k}_{1}s_{1}}^{\textcolor{black}{\dagger}}(t)
\psi_{\mathbf{k}_{1}\bar{s}_{1}}^{\textcolor{white}{\dagger}}(t)
\psi_{\mathbf{k}_{2}\bar{s}_{2}}^{\textcolor{black}{\dagger}}
\psi_{\mathbf{k}_{2}s_{2}}^{\textcolor{white}{\dagger}}
\right\rangle	, 
\end{split}
\end{equation}
including all the terms of the KB SCHF resummation, which includes the entire Bethe-Salpeter ladder series with all internal single-particle Green's functions dressed with the SCHF self-energy~\cite{baym1961conservation,PhysRev.127.1391}. These SCHF Green's functions are given by (see \S\ref{sect:Supp:Pert-Theory} in \cite{SupplementalMaterial}):
\begin{eqnarray}
\label{eq:Dressed-propagator}
G^{s_{1}s_{2}}_{\mathbf{k}}&(\omega) &= 
\begin{split}
\includegraphics[scale=0.6,page=1]{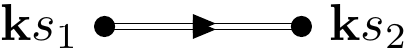}
\end{split} =
\frac{\delta_{s_{1}s_{2}}}{\omega-s_{1}(E_{\mathbf{k}}-i \eta)}.
\end{eqnarray}
and the $n$-th order Feynman diagram of this series is shown in Fig. \ref{fig:Dibujos}c. The zeroth order term of the series is:
\begin{equation}
\begin{split}
&{\chi^{(0)}}_{\mathbf{k}_{0}\mathbf{k}_{1}}^{s_{0}s_{1}}
(\omega) = 
-\delta_{\mathbf{k}_{0}\mathbf{k}_{1}}\delta_{s_{0}s_{1}}
\left(
\frac{\delta_{s_{0},1}-\delta_{s_{0},-1}}
{\omega - 2s_{0} (E_{\mathbf{k}_{0}} - i\eta ) }
\right).
\end{split} 
\end{equation}

An important property of this series, that can be readily obtained by integrating over internal intermediate frequencies, is that the intermediate Green's functions are all constrained to satisfy $s=s'$ which physically means that the intermediate pairs always have one electron in the conduction and the other in the valence band. 
This allows to cast the series as a matrix geometric series involving $\chi^{0}$ and $T$ of eq. \eqref{eq:Interacion-matrix} of the form:
\begin{equation*}
\begin{split}
&\chi(\omega) = 
{\chi^{0}}\!(\omega) + 
{\chi^{0}}\!(\omega)T \left( 
{\chi^{0}}\!(\omega) + 
{\chi^{0}}\!(\omega)T{\chi^{0}}\!(\omega) + 
\cdots 
\right),
\end{split}
\end{equation*}
and therefore the solution of the series has the form:
\begin{equation}
\chi^{-1}_{\mathbf{k}_{0}\mathbf{k}_{f}}(\omega) = 
-(\omega- i \eta) \tau^{z} \delta_{\mathbf{k}_{0}\mathbf{k}_{f}}
-H_{\mathbf{k}_{0}\mathbf{k}_{f}},
\end{equation}
where $H_{\mathbf{k}_{0}\mathbf{k}_{f}}$ is given in Eq. \eqref{eq:HP-Explicit-Hamiltonian} and $\tau^{z}$ is the diagonal Pauli matrix. 
The structure of this correlator is equal to the propagator of the HP bosons of the Hamiltonian \eqref{eq:HP-Explicit-Hamiltonian}. We therefore see that the exciton propagator has an identical effective Hamiltonian to the one obtained from the HP bosonic Hamiltonian in Eq. \eqref{eq:HP-Explicit-Hamiltonian}, demonstrating that the bosonized Hamiltonian is equivalent to self-consistent KB resummation of the particle-hole propagator.

\textit{\color{blue} Momentum space reparametrization}. 
So far we have imagined our system to have a finite size so that momenta belongs to a discrete lattice. However, it is convenient to perform a reparametrization that manifestly displays the symmetries of the thermodynamic limit. If we parametrize momentum space by a new coordinate $\mathbf{z}(\mathbf{k})$, we can trade our boson Hamiltonian by one in a different lattice given by:
\begin{equation}
\label{eq:Complete-Hamiltonian-Bogoliubov}
H_{HP} = \sum_{\mathbf{z},\mathbf{z'}}
B_{\mathbf{z }}^{\textcolor{black}{\dagger}}
H_{\mathbf{z }\mathbf{z'}}
B_{\mathbf{z'}}^{\textcolor{white}{\dagger}}.
\end{equation}



As detailed in~\cite{SupplementalMaterial}, in order to preserve the underlying microscopic normalization of the states, the boson operators and the Hamiltonian in the new lattice need to be rescaled as follows:
\begin{equation}
\label{eq:Lattice-transformations}
\begin{split}
B_{\mathbf{z}} &= J(\mathbf{z})
B_{\mathbf{k}},	\quad	
H_{\mathbf{z }\mathbf{z'}} = 
J(\mathbf{z})J(\mathbf{z'})
H_{\mathbf{k }\mathbf{k'}},
\end{split}
\end{equation}
where $J(\mathbf{z})=\sqrt{D(\mathbf{z})(\Delta z_1\Delta z_2)/(\Delta k_1\Delta k_2)}$, $\Delta k_i =2 \pi/L_i$, $\Delta z_i$ is the discretization unit of the new coordinate system and $D(\mathbf{z})$ is the Jacobian of the transformation. In particular, in order to exploit the emergent rotational invariance in the thermodynamic limit, we use the following polar parametrization $\mathbf{z}=(k,\phi)$:
\begin{equation}
\label{eq:Tangential-radius}
\begin{split}
    k_{m} &= \frac{\mathcal{K}}{\sqrt{2}} \tan^{2}(m \Delta \theta), 
    \quad \phi_{n}=n\Delta \phi,
\end{split}
\end{equation}
where $(k_m,\phi_n)$ are the polar coordinates of a given site in the polar momentum lattice depicted in Figs. \ref{fig:Dibujos}d and \ref{fig:Dibujos}e, $\mathcal{K}$ is the UV momentum scale, $\Delta \theta = (\sfrac{\pi}{2}) /(M+1)$, $\Delta \phi = 2\pi/(2L+1)$, and $n={0,...,2L}$, $m={1,...,M}$. 

The radial discretization we are choosing is denser at small $k$ and more dilute at large $k$. This is not crucial but allows faster numerical convergence at low energies. Notice also that we do not have a hard cutoff but the largest momentum apporaches infinity as $M \rightarrow \infty$. We have verified that the results we will describe are independent of the specific choice of the radial discretization once the grids become sufficiently dense \cite{SupplementalMaterial}.

Applying the transformation from Eq. \eqref{eq:Lattice-transformations} to the boson Hamiltonian from Eq. \eqref{eq:Complete-Hamiltonian-Bogoliubov} leads to the following decoupling into angular momentum channels:
\begin{equation}
\label{eq:mn-Hamiltonian}
\begin{split}
B_{m}^{\ell}  &= \sum_{\ell=-L}^{L}
e^{i \ell \phi_{n}}B_{mn}
,\quad	
H_{HP} = \sum_{m\ell}
B_{m}^{\ell\textcolor{black}{\dagger}}
H_{mm'}^{\ell}
B_{m'}^{\ell}
\end{split}
\end{equation}
Therefore the problem reduces to a set of bosons moving in an effective one dimensional radial space for each angular momentum channel which in general needs to be solved numerically.	

\textit{\color{blue} Optical Conductivity}. 
As a concrete application of our formalism we study the Coulomb interaction corrections to the optical conductivity of Dirac fermions. We follow the Kubo approach to compute the conductivity from the current-current correlator $\chi_{\mu\nu}(t)=i\Theta(t)A 
\left\langle
\left[j_{\mu}(t),j_{\nu}(0)
\right]
\right\rangle$. The total current operator carries $\mathbf{Q}=0$, so it can be represented {\textit{exactly}} within the effective spin-1/2 Hilbert space of Eq. \eqref{eq:Effective-Heisenberg-Hamiltonian} as follows:
\begin{equation}
\mathbf{j} = 
\frac{v}{A}
\sum_{\mathbf{k}}
\psi^{\textcolor{black}{\dagger}}_{\mathbf{k}\sigma_{1}}
\boldsymbol{\sigma}_{\sigma_{1}\sigma_{2}}
\psi^{\textcolor{white}{\dagger}}_{\mathbf{k}\sigma_{2}} = 
\frac{v}{A}
\sum_{\mathbf{k}}
\hat{\mathbf{s}}_{\mathbf{k}}
\end{equation}
Using the HP approximation for the spin operators, the current-current correlator then can be expressed as (see Eq. \eqref{eq:Supp:chi-Explicit} in \cite{SupplementalMaterial})
\begin{equation*}
\label{eq:Chi-phiphi-a-indices}
\begin{split}
\chi_{\varphi\varphi}(t) &= 
i\Theta(t)\frac{2v^{2}\Delta \theta}{(2\mathcal{K})^{2}(2\pi)}
\sum_{mm'}
\Scale[1.0]{
S_{m}
\left\langle
\left[
B_{m }^{1\textcolor{white}{\dagger}}(t),
B_{m'}^{1\textcolor{black}{\dagger}}
\right]
\right\rangle
S_{m'}
} , 
\end{split}
\end{equation*}
where $[B_{m}^{\ell\textcolor{black}{\dagger}}] = 
\begin{pmatrix}
b_{0}^{\ell\textcolor{black}{\dagger}}	&	\!\!\cdots\!\!	&	
b_{M}^{\ell\textcolor{black}{\dagger}}	&	\!\!
b_{0}^{\ell\textcolor{white}{\dagger}}	&	\!\!\cdots\!\!	&	
b_{M}^{\ell\textcolor{white}{\dagger}}
\end{pmatrix}$ and $[S_{m}] = 
\begin{pmatrix}
t_{0}	\;
\cdots	\;
t_{M}	\;	-
t_{0}	\;
\cdots	\;	-
t_{M}
\end{pmatrix}$ are scale factors with $t_{m}=\sqrt[3]{\tan(\theta_{m})}\sec(\theta_{m})$. Because the current transforms as a vector under rotations the calculation of the conductivity only requires solving the boson bilinear Hamiltonian of Eq. \eqref{eq:mn-Hamiltonian} for the angular momentum channel $\ell=1$. Then if the Hamiltonian for the $\ell=1$ angular momentum channel is diagonalized by a transformation of the form (see also \cite{vanHemmen1980note}):
\begin{equation}
B_{m}^{1\textcolor{white}{\dagger}} = 
\sum_{n}
R_{mn}^{\textcolor{white}{\dagger}}
D_{n}^{\textcolor{white}{\dagger}}, 
\;
H_{mm'}^{1} = 
\sum_{nn'}
R_{mn}^{\textcolor{white}{*}}
\Omega_{nn'}
R_{n'm'}^{\textcolor{black}{*}}
\end{equation}
where $\Omega_{nn'}=\mathrm{diag}\begin{pmatrix}
\omega_{0}
\cdots
\omega_{M}
-\!\omega_{0}
\!\cdots
\!-\omega_{M}
\end{pmatrix}$ is the diagonal matrix of the eigenvalues of Eq. \eqref{eq:HP-Hamiltonian}, the real part of the conductivity can be obtained from the following Lehmann-type representation (see Eq. \eqref{eq:Supp:OptCond-Analytic} in \cite{SupplementalMaterial}):
\begin{equation}
\label{eq:OptCond-Analytic}
\begin{split}
\sigma(\omega) &= 
\frac{v^{2}\Delta \theta}{(2\mathcal{K})^{2}}
\sum_{m}
\left|\sum_{n}R^{*}_{mn} 
S_{n}\right|^{2}
\frac{
\delta(\omega-\omega_{m})}
{\omega_{m}}.
\end{split}
\end{equation}	

\begin{figure}
\centering		
\includegraphics[scale=0.60]{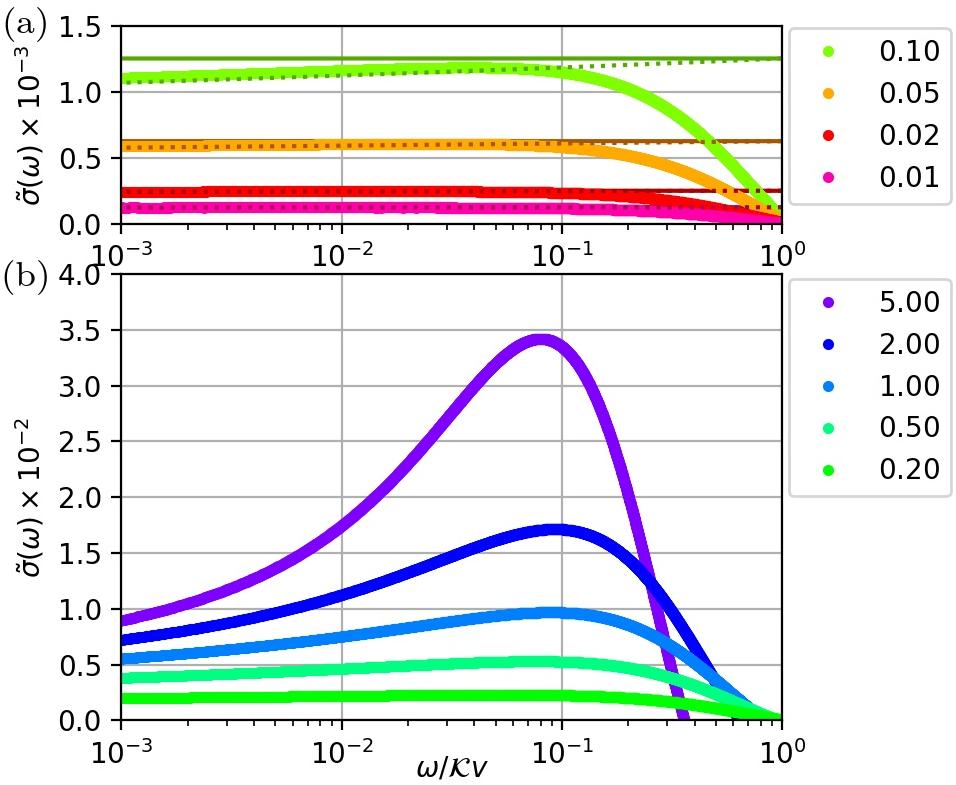}
\caption{(a) Conductivity at weak couplng. Thick: Numerical data. Dotted: RG correction $\tilde{\sigma}(\omega)$. Solid: perturbative correction from Ref. \citep{mishchenko2008minimal}. (b) Conductivity at strong coupling. }
\label{fig:Opt-Cond}
\end{figure}

We will now describe the results for the optical conductivity obtained by numerically diagonalizing the $\ell=1$ angular momentum bilinear Hamiltonian of Eq. \eqref{eq:mn-Hamiltonian} for the Coulomb interaction $V_{\mathbf{q}}=2\pi e^2/\epsilon |\mathbf{q}|$. Further details on the numerics can be found in \cite{SupplementalMaterial}.
To isolate the interaction corrections to $\sigma(\omega)$ we define:
\begin{equation}
\tilde{\sigma}(\omega) = \frac{\sigma(\omega)-\sigma_{0}}{\sigma_{0}}
\end{equation}
where $\sigma_0=e^2/16$ is the non-interacting conductivity of Dirac fermions.
The leading perturbative correction to this conductivity is expected to be of the form~\cite{mishchenko2008minimal}: $\tilde\sigma = C\alpha+\mathcal{O}(\alpha^2)$, with $C=(19-6\pi)/12$. We have been able to reproduce this perturbative correction numerically at small $\alpha$ as shown by solid horizontal lines in Fig. \ref{fig:Opt-Cond}a along with the full numerical result from Eq. \eqref{eq:OptCond-Analytic}. At larger values of $\alpha$, clear deviations from the leading perturbative result are seen in Fig. \ref{fig:Opt-Cond}b. One of the conspicuous deviations is a logarithmic decrease of the conductivivity at low frequencies (see \cite{SupplementalMaterial}). This logarithmic decrease can be explained by the logarithmic running of the coupling constant at small frequencies expected from the perturbative renormalization group (RG) analysis:
\begin{equation}
\label{eq:RG-OptCond}
\tilde{\sigma}(\omega) = \frac{C\alpha}{1+\frac{\alpha}{4}\ln\left( \frac{\mathcal{K}v}{\omega} \right)} 
\approx
C\alpha \left( 1 + \frac{\alpha}{4}\ln\left( \frac{\omega}{\mathcal{K}v} \right) \right)
\end{equation}
The predicted RG logarithmic correction is shown by a dotted line in Fig. \ref{fig:Opt-Cond}a which is in good agreement with the numerical implementation of Eq. \eqref{eq:RG-OptCond} at small $\alpha$. For larger values of $\alpha$ we see clear deviations from this leading RG perturbative result, as shown in Fig. 2b.

Nevertheless, as shown in Fig. \ref{fig:Opt-Cond}, even for a value of $\alpha$ as large as $\alpha=5$ the maximal deviation of the conductivity from the non-interacting value is only about 4\%. This indicates a resilience of conductivity of Dirac fermions to  interactions corrections even when non-pertubative effects are included, in agreement with experiments that have obtained values close to those of non-interacting fermions~\cite{li2008dirac, mak2008measurement, nair2008fine}. We even suspect that the interaction corrections in a full exact solution of the Hamiltonian in Eq. \eqref{eq:Main-Ham} would even be weaker than the corrections we obtained, because the RPA screening of the Coulomb interactions, roughly speaking, should lead to a reduction of the effective value of $\alpha\! \rightarrow \!\alpha_{\mathrm{RPA}} \approx \alpha/(1+\pi N \alpha/8)$, where $N$ is the total number of Dirac cones (e.g. $N=4$ for graphene~\cite{sodemann2012interaction}).



\textit{\color{blue} Discussion and Summary}.
We have developed a formalism that captures non-perturbatively the effects interactions on the continuum of $\mathbf{Q}=0$ particle-hole excitations of Dirac fermions. Our approach is constructed by projecting the full microscopic many-body Hamiltonian of Dirac fermions into the subspace of singly occupied momentum states, leading to an effective spin-1/2 Heisenberg-like model in a momentum lattice. This problem is subsequently reduced to a boson bilinear Hamiltonian by a standard Holstein-Primakoff transformation. We have provided a solid microscopic justification for this formalism by showing that it is equivalent to the Kadanoff-Baym resummation of the particle-hole propagator associated with the SCHF approximation to the single particle Green's function. This approximation is expected to capture the essential universal low energy properties of the semi-metallic phase that evolves adiabatically from Free fermions. We have applied this formalism to compute the Coulomb interaction corrections to the optical conductivity of Dirac fermions and found that it recovers the results of perturbative renormalization group at weak coupling~\cite{mishchenko2008minimal} and extended them to strong coupling. Remarkably, we have found that the Coulomb interaction corrections remain very weak ($\sim 4\%$) up to values of the effective fine structure constant $\alpha \sim 5$, in agreement with experiments in graphene that have measured a value of the optical conductivity that is consistent with the free electron theory~\cite{li2008dirac, mak2008measurement, nair2008fine}. Although our discussion has been restricted to 2D Dirac fermions, our approach can be naturally generalized to other multi-band semi-metals and higher dimensions, such as Weyl semimetals~\cite{armitage2018weyl} and novel nodal fermions~\cite{bradlyn2016beyond}, providing an interesting tool to capture non-perturbative effects of interactions on the correlation functions of $\mathbf{Q}=0$ operators of these phases.

\bibliography{Manuscript}
\bibliographystyle{ieeetr}

\clearpage			






\appendix

\setcounter{equation}{0}

\section*{SUPPLEMENTAL MATERIAL}

\section{Connection to perturbation theory}

\subsection{Band basis and interaction matrix}
\label{sect:Supp:Band-basis}
In this section, we provide details of the derivation of Eqs. \eqref{eq:HP-Hamiltonian} to \eqref{eq:Interacion-matrix} starting from Eq. \eqref{eq:Main-Ham}. We begin describing the transfomation from pseudo-spin basis onto band basis. In the band basis $s=\{+,-\}$ the kinetic term is ($e^{\pm i\phi} = \hat{k}_{x} \pm i \hat{k}_{y}$):
\begin{equation}
\begin{split}
\psi_{\mathbf{k }\sigma }^{\textcolor{black}{\dagger}}&
\left( \mathbf{k} \cdot 
\boldsymbol{\sigma}_{\sigma\sigma'} \right)
\psi_{\mathbf{k }\sigma'}^{\textcolor{white}{\dagger}}
 = 
\\ &= 
    k (
    e^{-i\phi}
    \psi_{\mathbf{k\uparrow  }}^{\textcolor{black}{\dagger}}
    \psi_{\mathbf{k\downarrow}}^{\textcolor{white}{\dagger}} + 
    e^{+i\phi}
    \psi_{\mathbf{k\downarrow}}^{\textcolor{black}{\dagger}}
    \psi_{\mathbf{k\uparrow  }}^{\textcolor{white}{\dagger}} 
    ).
\end{split}
\end{equation}
Band and pseudospin basis are related by:
\begin{equation}
\begin{split}
	\psi_{\mathbf{k}\sigma} &= 
	\sum_{s}
	\left\langle \sigma | \mathbf{k }s \right\rangle
	\psi_{\mathbf{k} s },	\\
\begin{pmatrix}
    \psi_{\mathbf{k\uparrow  }}	\\
    \psi_{\mathbf{k\downarrow}}	
\end{pmatrix} &= 
\frac{1}{\sqrt{2}}
\begin{pmatrix}
	 e^{- i\phi /2}	& 
	 e^{- i\phi /2}	\\  
	 e^{+ i\phi /2}	&
    -e^{+ i\phi /2}
\end{pmatrix}
\begin{pmatrix}
	\psi_{\mathbf{k+}}	\\
	\psi_{\mathbf{k-}}	
\end{pmatrix}.
\end{split}
\end{equation}
Fermion bilinears transform as
\begin{equation}
\begin{split}
\sum_{\sigma}
\psi_{\mathbf{k}_{1}\sigma}^{\textcolor{black}{\dagger}}
\psi_{\mathbf{k}_{2}\sigma}^{\textcolor{white}{\dagger}}
&= 
\sum_{s_{1}s_{2}}\!\!
\left\langle \mathbf{k}_{1}s_{1} | \mathbf{k}_{2}s_{2} \right\rangle \psi_{\mathbf{k}_{1}s_{1} }^{\textcolor{black}{\dagger}}
\psi_{\mathbf{k}_{2}s_{2} }^{\textcolor{white}{\dagger}},
\end{split}
\end{equation}
where
\begin{equation}
\begin{split}
& \left\langle \mathbf{k}_{1}s_{1} | \mathbf{k}_{2}s_{2} \right\rangle =
\begin{pmatrix}
     \cos\sfrac{\phi_{12}}{2}  &    i\sin\sfrac{\phi_{12}}{2} \\  
    i\sin\sfrac{\phi_{12}}{2}  &     \cos\sfrac{\phi_{12}}{2}
\end{pmatrix},
\end{split}
\end{equation}
where $\phi_i$ is the polar angle of $\mathbf{k}_i$, and  $\phi_{12}=\phi_1-\phi_2$. The part of the Hamiltonian in Eq. \eqref{eq:Main-Ham} that produces only $\mathbf{Q}=0$ inter-band transitions in the band basis is $
\psi_{\mathbf{k}_{1}s_{1} }^{\textcolor{black}{\dagger}}
\psi_{\mathbf{k}_{2}s_{2} }^{\textcolor{white}{\dagger}}
\psi_{\mathbf{k}_{2}\bar{s}_{2} }^{\textcolor{black}{\dagger}}
\psi_{\mathbf{k}_{1}\bar{s}_{1} }^{\textcolor{white}{\dagger}}
$. Therefore:
\begin{equation*}
\begin{split}
& \sum_{\sigma_{1} \sigma_{2}}
\psi_{\mathbf{k}_{1}\sigma_{2} }^{\textcolor{black}{\dagger}}
\psi_{\mathbf{k}_{2}\sigma_{2} }^{\textcolor{white}{\dagger}}
\psi_{\mathbf{k}_{2}\sigma_{1} }^{\textcolor{black}{\dagger}}
\psi_{\mathbf{k}_{1}\sigma_{1} }^{\textcolor{white}{\dagger}}
= \\ &= 	
\sum_{s_{1}s_{2}}\!
\left|\left\langle \mathbf{k}_{1}s_{1} | 
\mathbf{k}_{2}s_{2} \right\rangle \right|^{2}
\psi_{\mathbf{k}_{1}s_{1} }^{\textcolor{black}{\dagger}}
\psi_{\mathbf{k}_{2}s_{2} }^{\textcolor{white}{\dagger}}
\psi_{\mathbf{k}_{2}\bar{s}_{2} }^{\textcolor{black}{\dagger}}
\psi_{\mathbf{k}_{1}\bar{s}_{1} }^{\textcolor{white}{\dagger}}.
\end{split}
\end{equation*}
Consequently, the Hamiltonian in Eq. \eqref{eq:Main-Ham} of the main text projected onto the subspace of singly occupied momentum sites, defined in Eq. 
\eqref{eq:Effective-Heisenberg-Hamiltonian}, can be expressed as follows:
\begin{equation}
\label{eq:Supp:Projected-Hamiltonian}
\begin{split}
&\mathcal{P} H \mathcal{P} = 
v\sum_{\mathbf{k }}|\mathbf{k}|
\left( 
\psi_{\mathbf{k}+ }^{\textcolor{black}{\dagger}}
\psi_{\mathbf{k}+ }^{\textcolor{white}{\dagger}} - 
\psi_{\mathbf{k}- }^{\textcolor{black}{\dagger}}
\psi_{\mathbf{k}- }^{\textcolor{white}{\dagger}}
\right)
\\
&- 
\sum_{\mathbf{k}_{1}\neq \mathbf{k}_{2}} 
\sum_{s_{1} s_{2}} 
T_{\mathbf{k}_{1}\mathbf{k}_{2}}^{s_{1}s_{2}}
\psi_{\mathbf{k}_{1}s_{1} }^{\textcolor{black}{\dagger}}
\psi_{\mathbf{k}_{1}\bar{s}_{1} }^{\textcolor{white}{\dagger}}
\psi_{\mathbf{k}_{2}\bar{s}_{2} }^{\textcolor{black}{\dagger}}
\psi_{\mathbf{k}_{2}s_{2} }^{\textcolor{white}{\dagger}}
,
\end{split}
\end{equation}
with $T_{\mathbf{k}_{1}\mathbf{k}_{2}}^{s_{1}s_{2}}$ given by:
\begin{equation}
\label{eq:Supp:Interaction-matrix-band-basis}
T_{\mathbf{k}_{1}\mathbf{k}_{2}} = 
\frac{V_{\mathbf{k_{1}-k_{2}}}}{4A} 
\begin{pmatrix}
    1+\cos\phi_{12}  &    1-\cos\phi_{12} \\  
    1-\cos\phi_{12}  &    1+\cos\phi_{12}
\end{pmatrix}.
\end{equation}
The Hamiltonian of Eq. \eqref{eq:Supp:Projected-Hamiltonian} can then be expressed in terms of spin operators and leads to the Heisenberg-like model introduced in Eq. 
\eqref{eq:Effective-Heisenberg-Hamiltonian} of the main text.

\subsection{Holstein-Primakoff expansion}
\label{sect:Supp:Holstein-Primakoff}
We select the following spin basis 
\begin{equation}
\label{eq:Supp:Zeemann-basis}
\mathbf{s}_{\mathbf{k}} = -
s_{\mathbf{k}}^{z}\hat{\mathbf{k}} + 
s_{\mathbf{k}}^{x} \hat{\mathbf{z}} + 
s_{\mathbf{k}}^{y}\hat{\boldsymbol{\phi}},
\end{equation}
which diagonalizes the kinetic term. On this basis, the Hamiltonian can be expanded in a bosonic representation by means of the Holstein-Primakoff (HP) transformations ($S=\nicefrac{1}{2}$):
\begin{equation}
\label{eq:Holstein-Primakoff}
\begin{split}
s_{\mathbf{k }}^{z} &= 2\left( S - 
b_{\mathbf{k }}^{\textcolor{black}{\dagger}}
b_{\mathbf{k }}^{\textcolor{white}{\dagger}} \right) = 1 - 2 
b_{\mathbf{k }}^{\textcolor{black}{\dagger}}
b_{\mathbf{k }}^{\textcolor{white}{\dagger}}
, \\
s_{\mathbf{k }}^{x}  
&\approx \sqrt{2 S}\left( 
b_{\mathbf{k }}^{\textcolor{white}{\dagger}} + 
b_{\mathbf{k }}^{\textcolor{black}{\dagger}} \right) = 
b_{\mathbf{k }}^{\textcolor{white}{\dagger}} + 
b_{\mathbf{k }}^{\textcolor{black}{\dagger}}
, \\
i s_{\mathbf{k }}^{y}
&\approx \sqrt{2 S}\left( 
b_{\mathbf{k }}^{\textcolor{white}{\dagger}} - 
b_{\mathbf{k }}^{\textcolor{black}{\dagger}} \right) = 
b_{\mathbf{k }}^{\textcolor{white}{\dagger}} - 
b_{\mathbf{k }}^{\textcolor{black}{\dagger}}
.
\end{split}
\end{equation}
The term corresponding to the exchange coupling in Eq. \eqref{eq:Effective-Heisenberg-Hamiltonian} can be transformed into pairing and hopping terms of bosons up to bilinears:
\begin{equation}
\begin{split}
\mathbf{s}_{\mathbf{k }} \cdot \mathbf{s}_{\mathbf{k'}} &\approx 
\left(
1+
b_{\mathbf{k }}^{\textcolor{black}{\dagger}}
b_{\mathbf{k }}^{\textcolor{white}{\dagger}} + b_{\mathbf{k'}}^{\textcolor{black}{\dagger}}b_{\mathbf{k'}}^{\textcolor{white}{\dagger}}\right)  
\cos \phi_{\mathbf{k }\mathbf{k'}}
\\ &+ 
\left(
b_{\mathbf{k }}^{\textcolor{black}{\dagger}}
b_{\mathbf{k'}}^{\textcolor{white}{\dagger}} + 
b_{\mathbf{k }}^{\textcolor{white}{\dagger}}
b_{\mathbf{k'}}^{\textcolor{black}{\dagger}}\right) 
\left( 1 + \cos \phi_{\mathbf{k }\mathbf{k'}} \right) 
\\ &+ 
 \left(
b_{\mathbf{k }}^{\textcolor{black}{\dagger}}
b_{\mathbf{k'}}^{\textcolor{black}{\dagger}} + 
b_{\mathbf{k }}^{\textcolor{white}{\dagger}}
b_{\mathbf{k'}}^{\textcolor{white}{\dagger}}
\right) 
\left( 1 - \cos \phi_{\mathbf{k }\mathbf{k'}} \right).
\end{split}
\end{equation}

The resulting bosonic Hamiltonian after applying the HP transformations is:
\begin{equation}
\label{eq:Supp:HP-Hamiltonian}
\begin{split}
&H_{HP} = \sum_{\mathbf{k}} 2 v |\mathbf{k }| 
b_{\mathbf{k }}^{\textcolor{black}{\dagger}}
b_{\mathbf{k }}^{\textcolor{white}{\dagger}}
+
\sum_{\mathbf{k}\neq \mathbf{k'}} 
\frac{V_{\mathbf{k-k'}}}{A}
b_{\mathbf{k }}^{\textcolor{black}{\dagger}}
b_{\mathbf{k }}^{\textcolor{white}{\dagger}}
\cos \phi_{\mathbf{k }\mathbf{k'}}
\\ &+ 
\sum_{\mathbf{k}\neq \mathbf{k'}} 
\frac{V_{\mathbf{k-k'}}}{4A}
\left( 1 + \cos \phi_{\mathbf{k }\mathbf{k'}} \right) 
\left(	
b_{\mathbf{k }}^{\textcolor{black}{\dagger}}
b_{\mathbf{k'}}^{\textcolor{white}{\dagger}} + 
b_{\mathbf{k }}^{\textcolor{white}{\dagger}}
b_{\mathbf{k'}}^{\textcolor{black}{\dagger}}
\right) + 
\\ &+ 
\sum_{\mathbf{k}\neq \mathbf{k'}} 
\frac{V_{\mathbf{k-k'}}}{4A}
\left( 1 - \cos \phi_{\mathbf{k }\mathbf{k'}} \right) 
\left(	
b_{\mathbf{k }}^{\textcolor{black}{\dagger}}
b_{\mathbf{k'}}^{\textcolor{black}{\dagger}} + 
b_{\mathbf{k }}^{\textcolor{white}{\dagger}}
b_{\mathbf{k'}}^{\textcolor{white}{\dagger}}
\right). 
\end{split}
\end{equation}
The first line contains the kinetic and self-energy terms. The second line can be viewed as boson hopping terms in the momentum lattice. The third line can be viewed as pairing terms which change the number of bosons. Lastly, by using the Bogoliubov basis given by
\begin{equation}
\label{eq:Supp:Bogoliubov-basis}
B^{\dagger}_{\mathbf{k}} = 
\begin{pmatrix}
b^{\textcolor{black}{\dagger}}_{\mathbf{k}}	&	
b^{\textcolor{white}{\dagger}}_{\mathbf{k}}
\end{pmatrix},
\end{equation}
the Hamiltonian can be expressed as shown in Eq. \eqref{eq:HP-Hamiltonian} and \eqref{eq:HP-Explicit-Hamiltonian} of the main text. 

\subsection{Details of connection to perturbation theory}
\label{sect:Supp:Pert-Theory}
The propagator of the Dirac fermions without interactions is diagonal in the band basis and is given by:
\begin{eqnarray}
\begin{split}
G^{\textcolor{black}{(0)}}_{ss'}&(\omega,\mathbf{k}) \! = 
\includegraphics[scale=0.9]{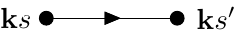}\! = \!
\frac{\delta_{ss'}}{\omega-s(v|\mathbf{k}|-i \eta)}.
\end{split}
\end{eqnarray}
Moreover, the Hartree-Fock self-energy of the fermions is:
\begin{eqnarray}
\begin{split}
\Sigma_{\mathbf{k }} = \!\!\!	
\begin{split}
\includegraphics[scale=0.9]{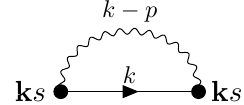}
\end{split} \!\!\! = 
\frac{1}{2A}\sum_{\mathbf{p}}V_{\mathbf{k-p}}\cos \phi_{\mathbf{kp}}.
\end{split}
\end{eqnarray}
We can then sum the Dyson series to get the dressed fermionic propagator:
\begin{eqnarray}
G^{\textcolor{white}{0}}_{ss'}&(\omega,\mathbf{k}) &= \!
\begin{split}
\includegraphics[scale=0.9]{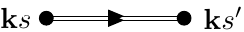}
\end{split}\! = \!\!
\frac{\delta_{ss'}}{\omega-s(E_{\mathbf{k}}-i \eta)},
\end{eqnarray}
where $E_{\mathbf{k}} = v|\mathbf{k}| + \Sigma_{\mathbf{k}}$. Equivalently in matrix notation we can write:
\begin{equation}
\label{eq:Supp:Dressed-Prop}
G^{\textcolor{black}{-1}}(\omega,\mathbf{k}) = 
\begin{pmatrix}
\omega - E_{\mathbf{k}} + i\eta &   0   \\
0   &   \omega + E_{\mathbf{k}} - i\eta
\end{pmatrix}.
\end{equation}

\begin{figure}
\centering
\includegraphics[scale=0.45]{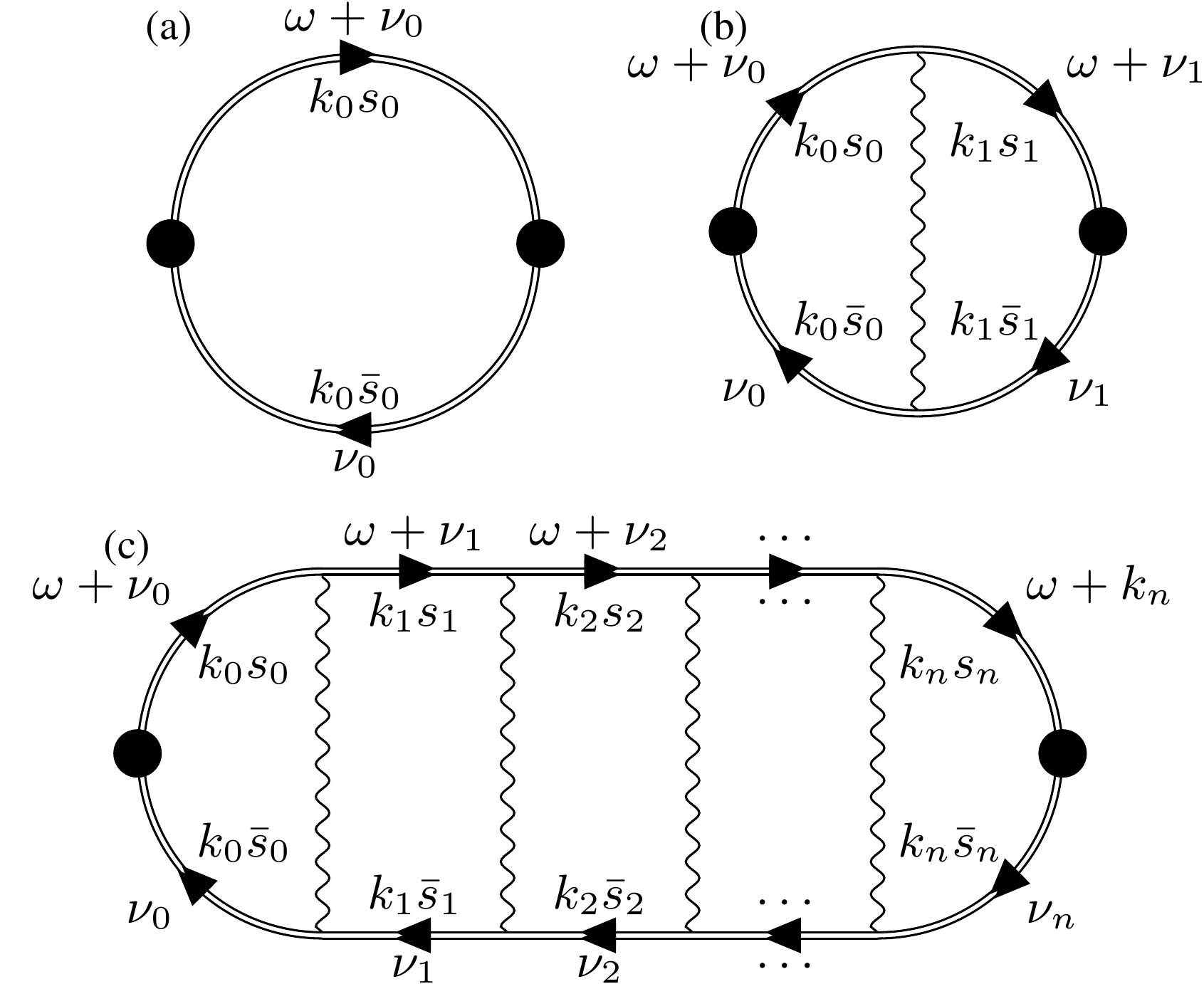}
\caption{Diagrams associated to the zeroth (a), first (b) and $n$-order (c) corrections of the Bethe-Salpeter ladder.}
\label{fig:Supp:Feynmann-diagrams-Bethe-Salpeter}
\end{figure}

Our goal is to compute the specific resummation of Feynman diagrams for the particle-hole propagator associated with the Kadanoff-Baym (KB) conserving approximation that results from the self-consistent Hartree-Fock (SCHF) approximation to the single particle Green's function. This resummation consists of the sum of the infinite series of the Bethe-Salpeter ladder for the particle-hole propagator with internal Green's functions dressed by the Hartree-Fock self energy from Eq. \eqref{eq:Supp:Dressed-Prop}. The series is depicted in Fig. \ref{fig:Supp:Feynmann-diagrams-Bethe-Salpeter}. The particle-hole propagator of interest is defined in Eq. \eqref{eq:Particle-hole-propagator}. The zeroth order non-interacting term of the series is given by:
\begin{eqnarray}
\label{eq:Supp:Corr-Func-4psi}
&& {\chi^{(0)}}^{ss'}_{\mathbf{k}}(\omega) =
- \int \dfrac{d\nu}{2\pi i}
G_{\mathbf{k}s }(\omega+\nu) 
G_{\mathbf{k}s'}(\nu) \\ &=&
- \int \dfrac{d\nu}{2\pi i}
\dfrac{1}{(\omega+\nu)+s'(E_{\mathbf{k}}-i\eta)}
\dfrac{1}{\nu+s(E_{\mathbf{k}}-i\eta)} \nonumber \\ &=&
- \delta_{s',\bar{s}}
\left(
\dfrac{\delta_{s,+}-\delta_{s,-}}
{\omega-(s-s')(E_{\mathbf{k}}-i\eta)}
\right) .\nonumber 
\end{eqnarray}
or expressed as a matrix in the band basis
\begin{equation}
\label{eq:Supp:Boson-propagator}
\chi^{(0)}_{\mathbf{k}_{0}}(\omega) = 
\begin{pmatrix}
\frac{-1}{\omega + (2 E_{\mathbf{k}_{0}} - i\eta ) } &   0   \\
0   &   \frac{1}{\omega - (2 E_{\mathbf{k}_{0}} - i\eta ) }
\end{pmatrix}.
\end{equation}

We will now illustrate the leading terms of the series involving the interaction matrix from Eq. \eqref{eq:Supp:Interaction-matrix-band-basis}. The diagrams involved at first order in the Bethe-Salpeter ladder are shown in the Fig. \ref{fig:Supp:Feynmann-diagrams-Bethe-Salpeter}b, and given by (we assume summation on any repeated index):
\begin{equation*}
\begin{split}
{\chi^{(1)}}_{\mathbf{k}_{0}\mathbf{k}_{\!f}}^{s_{0}s_{\!f}}
\!(\omega)\! &=
{\chi^{(0)}}_{\mathbf{k}_{0}\mathbf{k}_{\!f}}^{s_{0}s_{\!f}}
\!(\omega)\! + 
{\chi^{(0)}}_{\mathbf{k}_{0}}^{s_{0}}
\!(\omega)
{T}_{\mathbf{k}_{0}\mathbf{k}_{\!f}}^{s_{0}s_{\!f}}
{\chi^{(0)}}_{\mathbf{k}_{f}}^{s_{f}}
\!(\omega).\!
\end{split}
\end{equation*} 
Similarly the $n$-th term of the series, shown in Fig. \ref{fig:Supp:Feynmann-diagrams-Bethe-Salpeter}c, is given by
\begin{equation*}
\begin{split}
&{\chi^{(n)}}_{\mathbf{k}_{0}\mathbf{k}_{\!f}}^{s_{0}s_{\!f}}
\!(\omega)\! =
{\chi^{(0)}}_{\mathbf{k}_{0}\mathbf{k}_{\!f}}^{s_{0}s_{\!f}}
\!(\omega)\! + 
{\chi^{(0)}}_{\mathbf{k}_{0}}^{s_{0}}
\!(\omega)
{T}_{\mathbf{k}_{0}\mathbf{k}_{1}}^{s_{0}s_{\!f}}
{\chi^{(0)}}_{\mathbf{k}_{f}}^{s_{f}}
\!(\omega)\! \\ &+
{\chi^{(0)}}_{\mathbf{k}_{0}}^{s_{0}}
\!(\omega)
{T}_{\mathbf{k}_{0}\mathbf{k}_{1}}^{s_{0}s_{1}}
{\chi^{(0)}}_{\mathbf{k}_{1}}^{s_{1}}
\!\cdots 
{\chi^{(0)}}_{\mathbf{k}_{n\tiny{-}1}}^{s_{n\tiny{-}1}}	
{T}_{\mathbf{k}_{n\tiny{-}1}\mathbf{k}_{f}}^{s_{n\tiny{-}1}s_{f}}
{\chi^{(0)}}_{\mathbf{k}_{n}}^{s_{f}}	
\!(\omega).\!
\end{split}
\end{equation*}
The full summation can therefore be expressed as a geometric series:
\begin{equation*}
\begin{split}
&\chi(\omega) = {\chi^{0}}\!(\omega) + 
{\chi^{0}}\!(\omega)T{\chi^{0}}\!(\omega) + 
{\chi^{0}}\!(\omega)T{\chi^{0}}\!(\omega)T{\chi^{0}}\!(\omega) + 
\cdots \\ &= 
{\chi^{0}}\!(\omega) + 
{\chi^{0}}\!(\omega)T \left( 
{\chi^{0}}\!(\omega) + 
{\chi^{0}}\!(\omega)T{\chi^{0}}\!(\omega) + 
\cdots 
\right),
\end{split}
\end{equation*}
which correspond to a Dyson-like equation for the dressed particle-hole propagator ${\chi}(\omega)$:
\begin{equation*}
\begin{split}
\chi_{\mathbf{k}_{0}\mathbf{k}_{\!f}}^{s_{0}s_{\!f}}
\!(\omega)\! &=
{\chi^{(0)}}_{\mathbf{k}_{0}\mathbf{k}_{\!f}}^{s_{0}s_{\!f}}
\!(\omega)
\!+\!
{\chi^{(0)}}_{\mathbf{k}_{0}}^{s_{0}}\!(\omega)
T_{\mathbf{k}_{0}\mathbf{k}_{1}}^{s_{0}s_{1}}
{\chi}_{\mathbf{k}_{1}\mathbf{k}_{\!f}}^{s_{1}s_{\!f}}(\omega),
\end{split}
\end{equation*}
whose solution is given by:
\begin{equation}
\begin{split}
{\left(\chi^{-1}\right)}_{\mathbf{k}_{0}\mathbf{k}_{\!f}}^{s_{0}s_{\!f}}
\!(\omega)\! &=
{\delta}_{\mathbf{k}_{0}\mathbf{k}_{\!f}}^{s_{0}s_{\!f}}\!
{\left({\chi^{\Scale[0.5]{(}\Scale[0.6]{0}\Scale[0.5]{)}-1}}\right)}_{\mathbf{k}_{0}}^{s_{0}}\!(\omega)
\!+\!
T_{\mathbf{k}_{0}\mathbf{k}_{\!f}}^{s_{0}s_{\!f}}.
\end{split}
\end{equation}
Replacing the results from Eq. \eqref{eq:Supp:Boson-propagator} and \eqref{eq:Supp:Interaction-matrix-band-basis} we get:
\begin{equation}
\label{eq:Dressed-Boson-propagator}
\begin{split}
\chi^{-1}_{\mathbf{k}_{0}\mathbf{k}_{\!f}}(\omega) &= 
-\begin{pmatrix}
\omega + 2 E_{\mathbf{k}_{0}} - i\eta &   0   \\
0   &   \omega - 2 E_{\mathbf{k}_{0}} - i\eta  
\end{pmatrix} \\ &- 
\frac{V_{\mathbf{k}_{0}-\mathbf{k}_{\!f}}}{4A} 
\begin{pmatrix}
    1+\cos\phi_{0\!f}  &    1-\cos\phi_{0\!f} \\  
    1-\cos\phi_{0\!f}  &    1+\cos\phi_{0\!f}
\end{pmatrix}.
\end{split}
\end{equation}
or, by using the definition of the HP boson Hamiltonian in Eq. 
\eqref{eq:HP-Explicit-Hamiltonian} of the main text we get the final expression of the exciton propagator, given by
\begin{equation}
\chi^{-1}_{\mathbf{k}_{0}\mathbf{k}_{f}}(\omega) = 
-(\omega- i \eta) \tau^{z} \delta_{\mathbf{k}_{0}\mathbf{k}_{f}}
-H_{\mathbf{k}_{0}\mathbf{k}_{f}}.
\end{equation}

The structure of this correlator is identical to the propagator of the HP bosons of the Hamiltonian \eqref{eq:Supp:HP-Hamiltonian}. From the above, we can assert that the full resummation of the KB conserving approximation associated with SCHF is equivalent to solving the HP bilinear boson problem. 

\section{Momentum space reparametrization}
\subsection{General coordinate transformations on the continuum limit}
We begin by taking the continuum limit of the Hamiltonian in the Bogoliubov basis \eqref{eq:HP-Hamiltonian}, for this purpose it is convenient to define a rescaled Hamiltonian and boson creation operator as follows: 
\begin{equation}
\begin{split}
    B(\mathbf{k}) &\equiv \lim_{\Delta k\rightarrow 0} \frac{B_{\mathbf{k}}}{\sqrt{\Delta k_{x}\Delta k_{y}}},\\
    H(\mathbf{k },\mathbf{k'}) &\equiv \lim_{\Delta k \rightarrow 0}
    \frac{H_{\mathbf{kk'}}}{\Delta k_{x}\Delta k_{y}},
\end{split}
\end{equation}
where $\Delta k_{x,y}=2\pi/\sqrt{A}$ , A is the system area that we take to be a square. The discrete lattice of momenta with square symmetry is depicted in \ref{fig:Dibujos}d of the main text. The above re-definitions allow to obtain the following continuum commutation relations fo boson operators:
\begin{equation*}
    \left[
B(\mathbf{k }),
B^{\textcolor{black}{\dagger}}(\mathbf{k'})
\right] = \lim_{\Delta k\rightarrow 0}  \mathbb{I}\frac{\delta_{\mathbf{k k'}}}{(\Delta k)^2}
    = \mathbb{I}\delta^{2}(\mathbf{k-k'}),
\end{equation*}
where 
\begin{equation}
\mathbb{I} = 
\begin{pmatrix}
1	&	0	\\	0	&	-1	
\end{pmatrix}.
\end{equation}
With these rescalings we can convert the sums over momenta into continuum integrals, obtaining the continuum version of the boson Hamiltonian $H_{HP}$ from Eq. \eqref{eq:HP-Hamiltonian}:
\begin{equation*}
\begin{split}
\mathcal{H}_{HP} =& \lim_{\Delta k\rightarrow 0} \int \!\!
\frac{d^{2}k }{(\Delta k)^2}
\frac{d^{2}k'}{(\Delta k)^2}
B_{\mathbf{k }}^{\textcolor{black}{\dagger}}
H_{\mathbf{k } \mathbf{k'}}
B_{\mathbf{k'}}^{\textcolor{white}{\dagger}}\\
 =& \lim_{\Delta k\rightarrow 0} (\Delta k)^4 \!\! \int \!\!
\frac{d^{2}k }{(\Delta k)^2}
\frac{d^{2}k'}{(\Delta k)^2}
B_{\sigma }^{\textcolor{black}{\dagger}}(\mathbf{k })
H(\mathbf{k },\mathbf{k'})
B_{\sigma 
}^{\textcolor{white}{\dagger}}(\mathbf{k'})
\\
 =& \int \! d^{2}k d^{2}k'
\hat{B}_{\sigma }^{\textcolor{black}{\dagger}}(\mathbf{k })
H(\mathbf{k },\mathbf{k'})
\hat{B}_{\sigma 
}^{\textcolor{white}{\dagger}}(\mathbf{k'})
.
\end{split}
\end{equation*}

From this continuum Hamiltonian we can perfom a change of coordinates $\mathbf{k}(\mathbf{z})$ with Jacobian $D(\mathbf{z})=|\frac{\partial \mathbf{k}}{\partial \mathbf{z}}|$ with the following redefinitions:
\begin{equation}
\begin{split}
    B(\mathbf{z}) &= \sqrt{D(\mathbf{z})} 
    B(\mathbf{k}\Scale[0.9]{(\mathbf{z})}),
    \\
    H(\mathbf{z },\mathbf{z'})  &= \sqrt{D(\mathbf{z})D(\mathbf{z'})}
    H(\mathbf{k\Scale[0.9]{(\mathbf{z })}},\mathbf{k\Scale[0.9]{(\mathbf{z'})}}),
\end{split}
\end{equation}
whose purpose is to mantain the same form of the commutation relations and the Hamiltonian as follows:
\begin{equation*}
\begin{split}
&\left[
B(\mathbf{z }),B^{\textcolor{black}{\dagger}}(\mathbf{z'}) \right] =
\mathbb{I}\delta^{2}(\mathbf{z-z'}),
\\
&\mathcal{H}_{HP}
 = \int d^{2}z d^{2}z'
\hat{B}_{\sigma }^{\textcolor{black}{\dagger}}(\mathbf{z })
H(\mathbf{z },\mathbf{z'})
\hat{B}_{\sigma 
}^{\textcolor{white}{\dagger}}(\mathbf{z'})
.
\end{split}
\end{equation*}

Lastly, on the new coordinate system, we proceed to re-discretize the expressions, as follows:
\begin{equation}
\begin{split}
    B_{\mathbf{z}} &\leftarrow \sqrt{\Delta z_{1}\Delta z_{2}} 
    B(\mathbf{z}),
    \\
    H_{\mathbf{z },\mathbf{z'}} &\leftarrow \Delta z_{1}\Delta z_{2} \,
    H(\mathbf{z},\mathbf{z'}),
\end{split}
\end{equation}
that yield the new discrete commutation relations and Hamiltonian
\begin{equation*}
\begin{split}
\left[
B^{\textcolor{white}{\dagger}}_{\mathbf{z }},
B^{\textcolor{black}{\dagger}}_{\mathbf{z'}}
\right] &= 
\mathbb{I} \delta_{\mathbf{zz'}} 
\leftarrow \mathbb{I} 
\Delta k_{1} \Delta k_{2}  
\delta^{2}(\mathbf{z-z'})
\end{split}
\end{equation*}
\begin{equation}
\label{eq:Supp:Complete-Hamiltonian-Bogoliubov}
H_{HP} = \sum_{\mathbf{z},\mathbf{z'}}
B_{\mathbf{z }}^{\textcolor{black}{\dagger}}
H_{\mathbf{z }\mathbf{z'}}
B_{\mathbf{z'}}^{\textcolor{white}{\dagger}}.
\end{equation}
Therefore, in summary, the relation between operators and the Hamiltonian matrix in the new lattice defined by the discretization of the coordinates $\mathbf{z}(\mathbf{k})$, with the original operators and Hamiltonian of the square lattice is:
\begin{equation}
\label{eq:Supp:Lattice-transformations}
\begin{split}
B_{\mathbf{z}} &= \sqrt{D(\mathbf{z})\frac{\Delta z_{1}\Delta z_{2}}{\Delta k_{x}\Delta k_{y}}} 
B_{\mathbf{k}}  \\
H_{\mathbf{z }\mathbf{z'}} &= \sqrt{D(\mathbf{z })D(\mathbf{z'})}
\frac{\Delta z_{1}\Delta z_{2}}{\Delta k_{x}\Delta k_{y}}
H_{\mathbf{k }\mathbf{k'}}
\end{split}
\end{equation}
The idea is that the Hamiltonian $H_{HP}$ in Eq. \eqref{eq:Supp:Complete-Hamiltonian-Bogoliubov} will produce the same physical results as the one in the square lattice in Eq. \eqref{eq:HP-Hamiltonian} of the main text in the thermodynamic limit.

\subsection{Polar re-discretization}
We choose $\mathbf{z}=(k,\phi)$ where $k$ is the radius of the momentum vector and $\phi$ its polar angle. We will discretize the radial direction in a non-uniform way, to make it denser at small momenta and more dilute at large momenta. We have checked numerically that the precise form of the discretization is not crucial, but the choice we are making produces faster convergence to the thermodynamic limit. Therefore we choose the radius to be:
\begin{equation}
\label{eq:Supp:Tangential-radius}
\begin{split}
    k(\theta) &= \frac{\mathcal{K}}{\sqrt{2}} \tan^{2}(\theta),
\end{split}
\end{equation}
where $\mathcal{K}$ is a UV momentum scale, and $\theta \in (0,\pi/2)$ is another parameter labeling the radial coordinate that we will choose to be uniformly discretized. The corresponding Jacobian for this parametrization is:
\begin{equation}
\label{eq:Supp:Jacobian}
\begin{split}
    D(\theta) &= k\frac{dk}{d\theta} = \mathcal{K}^2 \frac{\tan^{3}(\theta)}{\cos^{2}(\theta)}.
\end{split}
\end{equation}
We choose $\theta$ and $\phi$ to be uniformly discretized as follows:
\begin{equation}
\label{eq:Supp:Discretization-of-angles}
\begin{split}
\theta_{m} &= m \Delta\theta, \quad m \in \{1,\cdots ,M\}, \\
  \phi_{n} &= n \Delta\phi  , \quad \:n \,\in \{0,\cdots ,2L\},
\end{split}
\end{equation}
where
\begin{equation}
\label{eq:Discrete-angle-units}
\Delta \theta = \frac{\sfrac{\pi}{2}}{M+1}, \quad
\Delta \phi = \frac{2\pi}{2L + 1}.
\end{equation}

After replacing \eqref{eq:Supp:Tangential-radius} and  \eqref{eq:Supp:Jacobian} into \eqref{eq:Supp:Lattice-transformations} we get the expression for $B_{\mathbf{k}}$ and $H_{\mathbf{k k'}}$ in the polar lattice
\begin{equation}
\label{eq:Supp:Lattice-transformations-1}
\begin{split}
B_{m}^{n\textcolor{white}{\dagger}} &=
\frac{\mathcal{K}}{2\pi}\sqrt{
A\Delta \theta \Delta \phi}
t_{m}
B_{\mathbf{k}_{mn}}^{\textcolor{black}{\dagger}},  \\
H_{mm'}^{nn'} &= 
\frac{\mathcal{K}^{2}}{(2\pi)^{2}}
A \Delta \theta \Delta \phi 
t_{m}t_{m'}
H_{\mathbf{k}_{mn}\mathbf{k}_{m'n'}},
\end{split}
\end{equation}
where $t(\theta_{m})=\sqrt{\tan(\theta_{m})}\sec(\theta_{m})$ and $\mathbf{k}_{mn}=\mathbf{k}(\theta_{m},\phi_{n})$. Finally, the whole Hamiltonian is
\begin{equation}
H_{HP} = \sum_{mn}\sum_{m'n'}
B_{m }^{n \textcolor{black}{\dagger}}
H_{mm}^{nn'}
B_{m'}^{n'\textcolor{white}{\dagger}}.
\end{equation}

\subsection{Angular momentum channels}
Because the Hamiltonian matrix $H_{\mathbf{k k'}}$ that enters into the Hamiltonina $H_{HP}$ in Eq. \eqref{eq:HP-Hamiltonian} of the main text only depends on the difference between the polar angles $\phi-\phi'$ we have conservation of the angular momentum $l$ of the bosons. Consequently, we perform Fourier transforms on the polar angles for the fields $B_{mn}$ and the matrix $H_{mn,m'n'}$
\begin{equation}
\begin{split}
B_{m}^{n\textcolor{white}{\dagger}} &= 
\frac{1}{\sqrt{2L+1}}\sum_{\ell=-L}^{L}
e^{-i \ell \phi_{n}}B_{m}^{\ell} ,
\\
H_{mm'}^{nn'} &= \sum_{\ell=-L}^{L}
e^{-i \ell (\phi_{n }-\phi_{n'})}
H_{mm'}^{\ell},
\end{split}
\end{equation}
such that the total Bogoliubov Hamiltonian decomposes into a direct sum for different angular mommentum channels, as follows:
\begin{equation}
H_{HP} = \sum_{mm'\ell}
B_{m}^{\ell\textcolor{black}{\dagger}}
H_{mm'}^{\ell}
B_{m'}^{\ell}.
\end{equation}

\section{Optical Conductivity}
\label{sect:Conductivity}

\subsection{Current density}
The current density operator is given by the spinor bilinear
\begin{equation}
\mathbf{j} = \frac{1}{A}\sum_{\mathbf{k}}v
\Psi_{\mathbf{k}}^{\textcolor{black}{\dagger}}
\boldsymbol{\sigma}
\Psi_{\mathbf{k}}^{\textcolor{black}{\dagger}}
 = \frac{1}{A}\sum_{\mathbf{k}}v
\hat{\mathbf{s}}_{\mathbf{k}}.
\end{equation}
After applying the Holstein-Primakoff expansion for the spins, this current can be expanded in terms of bosonic terms, and the leading expression is linear in the bosons and given by: 
\begin{equation}
\begin{split}
\mathbf{j} &\approx 	
v \sum_{\mathbf{k}}
s^{y}_{\mathbf{k}} 
\hat{\boldsymbol{\phi}} 
= 
v \sum_{\mathbf{k}}
i(b_{\mathbf{k }}^{\textcolor{black}{\dagger}}-
 b_{\mathbf{k }}^{\textcolor{white}{\dagger}})
\hat{\boldsymbol{\phi}} 
= 
v \sum_{\mathbf{k}}
i 
B_{\mathbf{k }}^{\textcolor{black}{\dagger}}\mathbf{I}
\hat{\boldsymbol{\phi}} ,
\end{split}
\end{equation}
where we also did the HP transformations \eqref{eq:Holstein-Primakoff} and chose the Bogoliubov basis \eqref{eq:Supp:Bogoliubov-basis} with the vector $\mathbf{I}=\mathrm{diag}(\mathbb{I})=\mathrm{(1,-1)}^{\mathrm{T}}$. Then, the optical conductivity is obtained as the imaginary part of the susceptibility
\begin{equation}
\chi_{\mu\nu}(t)=i\Theta(t)A 
\left\langle
\left[j_{\mu}(t),j_{\nu}(0)
\right]
\right\rangle.
\end{equation}
Without loss of generality, we choose the $x$-component of the current density:
\begin{equation}
\begin{split}
j_{x} &= 
v \sum_{\mathbf{k}}
i \mathbf{I}^{\mathrm{T}}
B_{\mathbf{k }}^{\textcolor{white}{\dagger}}
\sin \phi_{\mathbf{k}}.
\end{split}
\end{equation}

We can now apply the lattice transformations on $j_{x\phi}$
\begin{eqnarray}
j_{x\phi} &=& 	
i\frac{v \mathcal{K}\sqrt{\Delta \theta \Delta \phi}}{2\pi \sqrt{A}}
\sum_{mn}
S_{m} B_{m}^{n\textcolor{white}{\dagger}}
\sin \phi_{n}
,
\end{eqnarray}
where $m \in \{1,\cdots ,2M\}$, and the Jacobian of the transformation has been combined with the vector $\mathbf{I}$ as follows
\begin{equation}
[S_{m}] = 
\begin{pmatrix}
t_{0}	\;
\cdots	\;
t_{M}	\;	-
t_{0}	\;
\cdots	\;	-
t_{M}
\end{pmatrix}.
\end{equation}
with $t_{m}=\sqrt[3]{\tan(\theta_{m})}\sec(\theta_{m})$. Then, the Fourier transform is done, obtaining
\begin{equation}
\label{eq:Supp:xphi-Current}
j_{x} = 	
i\frac{v \mathcal{K}\sqrt{\Delta \theta}}{2 \sqrt{2\pi A}}
\sum_{m}
S_{m} (
B_{m}^{ 1\textcolor{white}{\dagger}}	-	
B_{m}^{-1\textcolor{white}{\dagger}} )
.
\end{equation}

The corresponding susceptibility is given by
\begin{eqnarray}
\label{eq:Supp:chi-Explicit}
\chi(t)&=&i\Theta(t)	
\frac{v^{2}\mathcal{K}^{2}\Delta \theta}{8\pi }\times \\ &\times	&
\sum_{mm'}
\left(
S_{m }
\left\langle
\left[
B_{m }^{ 1 }(t),
B_{m'}^{ 1 \textcolor{black}{\dagger}}(0)
\right]
\right\rangle 
S_{m'}
\right. \nonumber \\
&&+	
S_{m }
\left.
\left\langle
\left[
B_{m }^{-1}(t),
B_{m'}^{-1\textcolor{black}{\dagger}}(0)
\right]
\right\rangle
S_{m'}
\right).\nonumber
\end{eqnarray}
Inversion symmetry guarantees that $\ell=1$ contributes the same as $\ell=-1$, so that
\begin{equation}
\begin{split}
\label{eq:Supp:Last-chi-l=1}
\chi(t)=i & \Theta(t)
\frac{v^{2}\mathcal{K}^{2}\Delta \theta}{8\pi } 
\times  	\\  	&\times
\sum_{mm'}
S_{m }
\left\langle	
\left[
B_{m}^{     1 \textcolor{white}{\dagger}}(t),
B_{m}^{     1 \textcolor{black}{\dagger}}(0)
\right]
\right\rangle 
S_{m'}
.
\end{split}
\end{equation}

\subsection{Representation of the optical conductivity}
Let us assume that we diagonalize the 
Hamiltonian from Eq.(15) for the l=1 angular momentum channel via a Bogoliubov transformation, expressed as follows:
\begin{eqnarray}
B_{m}^{1\textcolor{white}{\dagger}} &=& 
\sum_{n}
R_{mn}^{\textcolor{white}{\dagger}}
D_{n}^{\textcolor{white}{\dagger}}, \\
\;
H_{mm'}^{1} &=& 
\sum_{nn'}
R_{mn}^{\textcolor{white}{*}}
\Omega_{nn'}
R_{n'm'}^{\textcolor{black}{*}},
\end{eqnarray}
where $\Omega_{nn'}=\mathrm{diag}\begin{pmatrix}
\omega_{0}\;
\cdots
\omega_{M}\;
-\omega_{0}\;
\cdots
-\omega_{M}
\end{pmatrix}$ is the diagonal matrix of the eigenvalues of the $\ell=1$ block of the HP Hamiltonian \cite{vanHemmen1980note}. Replacing such transformations in Eq. \eqref{eq:Supp:Last-chi-l=1} we get
\begin{eqnarray*}
&&\chi(t)=i\Theta(t)	
\frac{v^{2}\mathcal{K}^{2}\Delta \theta}{8\pi }\times \\
&\times &\sum_{mn}
e^{-i\Omega_{nn}t}
S_{m }R_{m n }^{\textcolor{white}{*}}
\left\langle	
\left[
D_{n }^{     1 \textcolor{white}{\dagger}},
D_{n'}^{     1 \textcolor{black}{\dagger}}
\right]
\right\rangle 
R_{n'm'}^{\textcolor{black}{*}}S_{m'}
.\nonumber
\end{eqnarray*}
which, because of $\left\langle	
\left[
D_{n }^{     1 \textcolor{white}{\dagger}},
D_{n'}^{     1 \textcolor{black}{\dagger}}
\right]
\right\rangle = \mathbb{I}_{nn'}$, yields 
\begin{eqnarray*}
&&\chi(t)=i\Theta(t)	
\frac{v^{2}\mathcal{K}^{2}\Delta \theta}{8\pi }\sum_{mn}
e^{-i\Omega_{nn}t}
S_{m }R_{m n }^{\textcolor{white}{*}}
\mathbb{I}_{nn'}
R_{n'm'}^{\textcolor{black}{*}}S_{m'}
.\nonumber
\end{eqnarray*}
Then, we take the Fourier transform of $\chi(t)$ to get the frequency-dependent susceptibility
\begin{eqnarray*}
\label{eq:Supp:Chi-Fourier}
&&\chi(\omega)=
\frac{v^{2}\mathcal{K}^{2}\Delta \theta}{8\pi }
\sum_{mn}
\frac{S_{m }R_{m n }^{\textcolor{white}{*}}
\mathbb{I}_{nn'}
R_{n'm'}^{\textcolor{black}{*}}S_{m'}}
{\omega - \Omega_{nn}+i\eta}
.\nonumber
\end{eqnarray*}

Finally, we take the imaginary part of Eq. \eqref{eq:Supp:Chi-Fourier} to get the optical conductivity depending on frequency as the following Lehmann-type representation:
\begin{equation}
\label{eq:Supp:OptCond-Analytic}
\begin{split}
\sigma(\omega) &= 
-\frac{e^2}{\omega}
\mathrm{Im}[\chi(\omega)] \\ &=
\frac{v^{2}\mathcal{K}^{2}\Delta \theta}{4}
\sum_{m}
\left|\sum_{n}R^{*}_{mn} 
S_{n}\right|^{2}
\frac{
\delta(\omega-\omega_{m})}
{\omega_{m}}.
\end{split}
\end{equation}	

\section{Numerical results}
As with any Lehmann representation of a conductivity, Eq.(18) is understood as a sequence of Dirac delta functions that approaches a continuous function in the thermodynamic limit. To obtain such continuous function it is useful to replace the Delta delta functions by a distribution that integrates to 1 but has a width that is larger than the finite size energy level spacing. A particularly convenient choice is to replace the Delta functions by retangular distributions with width $\Delta\omega_{m} = \omega_{m} - \omega_{m-1}$ and height $1/\Delta\omega_{m}$, between two adjacent energy levels $\omega_{m}$ and $\omega_{m-1}$, where we have assumed that the energy levels are ordered as $\omega_m > \omega_ {m - 1}$ . With this, we find that the conductivity can be approximated as:
\begin{equation}
\label{eq:Supp:OptCond-Numerical}
\begin{split}
\sigma(\omega_{m}) &= 
\frac{v^{2}\mathcal{K}^{2}\Delta \theta}{4}
\frac{
\left|\sum_{n}R^{*}_{mn} 
S_{n}\right|^{2}
}
{\omega_{m}(\omega_{m}-\omega_{m-1})}.
\end{split}
\end{equation}
All the plots presented for the optical conductivity correspond to the relative optical conductivity by substracting $\sigma_{0}=e^2/16$ and dividing by $\sigma_{0}$:
\begin{equation}
\label{eq:Supp:Sigma-tilde}
\tilde{\sigma}(\omega) = \frac{\sigma(\omega)-\sigma_{0}}{\sigma_{0}}.
\end{equation}	

To perform numerical calculations we have solved the Bogoliubov Hamiltonian of bosons from Eqs. \eqref{eq:HP-Hamiltonian} to \eqref{eq:Interacion-matrix} of the main text using a Coulomb interaction that has an explicit short distance (UV) and large distance (IR) regularization of the form:
\begin{equation}
\label{eq:Supp:Coulomb-Potential}
V_{\mathbf{q}} = \frac{
e^{-|\mathbf{q}|/\mathcal{K}} - 
e^{-|\mathbf{q}|/\mathcal{K}_{\mathrm{IR}}}}
{|\mathbf{q}|}
\end{equation}
where $\mathcal{K}$ is the large momentum cutoff and $\mathcal{K}_{\mathrm{IR}}$ is the small momentum cutoff. Physically $\mathcal{K}$ is of the order of the inverse lattice spacing and $\mathcal{K}_{\mathrm{IR}}$ can be literally viewed as controlled by the inverse distance to a metallic plane where image charges are produced. Throughout the paper we have used the UV cutoff $\mathcal{K}$ as the unit of momentum and $v\mathcal{K}$ as the unit of frequency. Although our model could have been used to study this physically sensible situation, we have focused on results that are universal and independent of these cutoffs. To do so we have only extracted information that numerically remains invariant as the cutoffs are respectively sent to infinity and zero. We describe the details of this procedure in the remainder of this supplementary section.

\subsection{Discretization size dependence}
Because the absolute correction to the conductivity remains small even up to larger values of $\alpha \sim 5$, it is important to ensure that our results converge as the size of discretization grid grows. We will now describe details of the dependence of the numerically computed conductivity on the size of the discretization grid.

\begin{figure}[h]
\centering	
\includegraphics[scale=0.8]{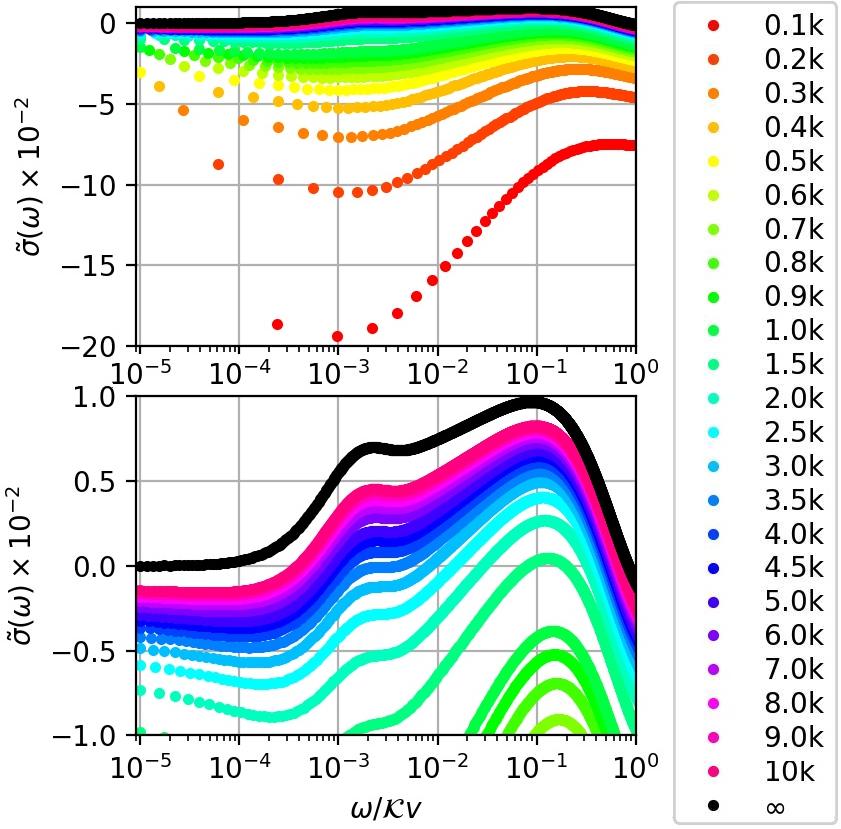}%
\caption{Optical conductivity $\tilde\sigma(\omega)$ vs. system size or discretization of the $k$ axis with $\sfrac{\mathcal{K}_{\mathrm{IR}}}{\mathcal{K}}=10^4$ and $\alpha=1$.}
\label{fig:System-size-dependence}
\end{figure}

To recapitulate, the numerical problem for the claculation of the conductivity reduces to the solution of Bogoliubov Hamiltonian of bosons for the l=1 channel. This Hamiltonian only has a nontrivial radial momentum coordinate, which we have discretized as follows:
\begin{equation}
k_{m} = \frac{\mathcal{K}}{\sqrt{2}}\tan^{2}(\theta_{m})
 = \frac{\mathcal{K}}{\sqrt{2}}
 \tan^{2}\left(\frac{\sfrac{\pi}{2}}{M+1}\right).
\end{equation}
$M$ is the integer labeling the total number of radial discrete momenta we include in the Hamiltonian, and we have taking it to range from $M=10^{2}$ to $10^{4}$. Figure \ref{fig:System-size-dependence} illustrates the behavior of the conductivity as a function of $M$ for $\alpha=1$.

The extrapolation of large $M$ is done using the values of $10^{3} \leq M \leq 10^{4}$, by fitting a linear function that depends on $1/M$. To make sure that the extrapolation does not change if we choose a different $6\times 10^{3} \leq M \leq 10^{4}$, and have verified that both extrapolations produce curves that lie on top of each other with the essentially the same values.

\subsection{Dependence on the IR cutoff}
\begin{figure}[h]
\centering
\includegraphics[scale=0.7]{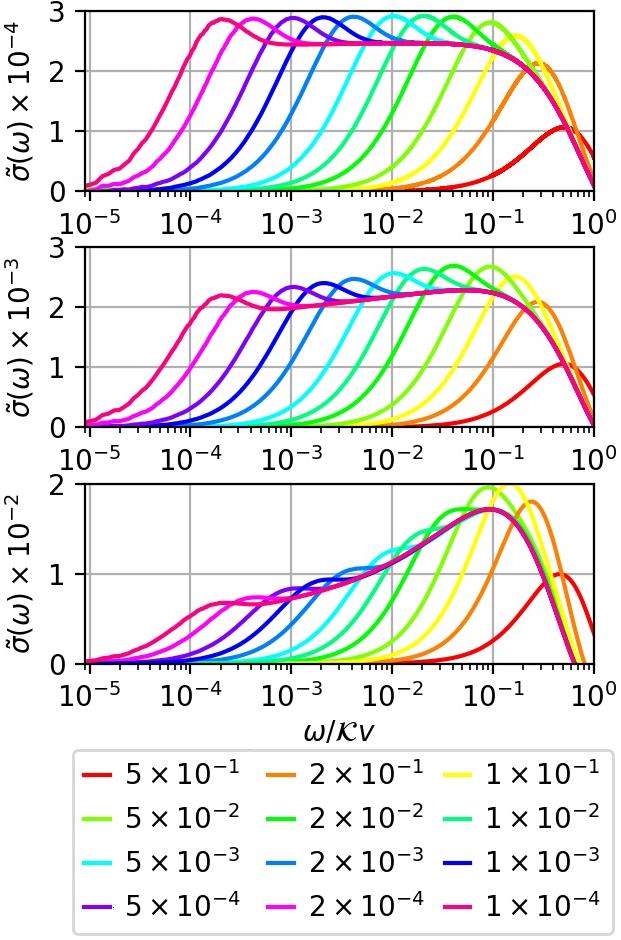}
\caption{The top panel corresponds to $\alpha = 0.02$, the middle to $\alpha = 0.2$, and the bottom to $\alpha = 2$. The lower inset shows the color convention for different choices of IR cutoff.}
\label{fig:IR-Cutoff-dependence}
\end{figure}

In Fig. \ref{fig:IR-Cutoff-dependence} we plot the behavior of the conductivity extrapolated to $M \rightarrow \infty$ for different values of the IR cutoff (each panel is for a different fixed value of $\alpha$). We see in Fig. \ref{fig:IR-Cutoff-dependence}, that at extremely low frequencies the conductivity has a bump followed by drop that changes with the value of the IR cutoff. Therefore this bump and the drop which occur at very low frequencies is a consequence of the IR cutoff which models screening of the Coulomb at long distances as captured by Eq. \eqref{eq:Supp:Sigma-tilde}.

Therefore, although this low frequency behavior is not completely unphysical as it models the behavior of the optical conductivity in the presence of a perfect metallic screening gate, it is not part of the universal behavior of the ideal unscreened Coulomb interaction that we are interested in. However, Fig. \ref{fig:IR-Cutoff-dependence} demonstrates clearly that the conductivities follow a universal curve because they agree perfectly at higher frequencies. Eventually the conductivity escapes from this universal curve at low frequencies when the IR cutoff becomes important. Therefore, we conclude from Fig. \ref{fig:IR-Cutoff-dependence}, that if we see that two curves with different IR cutoff overlap until some low frequency, and below this frequency they start to deviate from each other, the behavior for frequencies above this frequency at which they deviate is universal and represents the behavior for the ideal ideal Coulomb interaction problem without any IR cutoff. The results presented in the main text correspond to frequencies ranges where we observe independence of the IR cutoff, and where we are confident that we are simulating the ideal behavior of the unscreened Coulomb interaction.

\subsection{Non-perturbative effects of large $\alpha$ on the optical conductivity}
\begin{figure}[b]
\centering		
\includegraphics[scale=0.67]{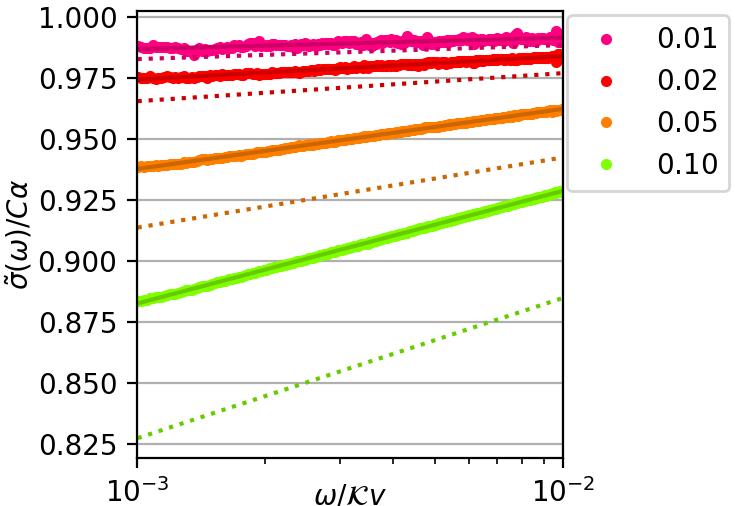}
\caption{Numerical calculation of the conductivity (color lines) and the the expected value from the leading order perturbative RG (dotted lines). The logarithmic running of the coupling constant leads to a visible linear logarithmic drift of the conductivity at weak coupling.}
\label{fig:Linear-regression}
\end{figure}

According to the perturbative RG result described in Ref. \cite{mishchenko2008minimal}, the optical conductivity is expected to have the following behavior at small frequencies and small $\alpha$:
\begin{equation}
\label{eq:Mishchenkos-expanded}
\begin{split}
\tilde{\sigma}(\omega) = 
\frac{C \alpha}{1 + \frac{\alpha}{4}\ln\left(\frac{\mathcal{K}v}{\omega}\right)} 
\approx 
C\alpha \left( 1 + \frac{\alpha}{4}\ln\left(\frac{\omega}{\mathcal{K}v}\right) \right),
\end{split}
\end{equation}
where $C=\frac{19-6\pi}{12}$. In the second line of this equation we have expanded the denominator in $\alpha$ to get the leading logarithmic correction to the conductivity. Other logarithmic corrections are expected to contain higher powers of $\alpha$. We have indeed observed such a weak logarithmic drift of the conductivity with frequency at small values of $\alpha$, as depicted in Fig. \ref{fig:Linear-regression}. By fitting the conductivity with a logarithmic dependence:
\begin{equation}
\tilde{\sigma}_{\mathrm{lin}}(\omega) = \tilde{\sigma}_{0} + \tilde{\sigma}_{1}\ln\left(\frac{\omega}{\mathcal{K}v}\right),
\end{equation}
we obtained the coefficients which are listed in Table \ref{tab:Linear-regression}. 

\begin{table}[h]
\centering	 	
\begin{tabular}{cccccc}
$\alpha$	&	$C\alpha\times 10^{-4}$	&	$C\alpha^{2}/4\times 10^{-7}$	&	$\tilde{\sigma}_{0}\times 10^{-4}$	&	$\tilde{\sigma}_{1}\times 10^{-7}$	\\
\hline \hline	
$0.01$	&	$1.254$	&	$3.134$	&	$1.255$	&	$2.552$	\\
$0.02$	&	$2.507$	&	$12.53$	&	$2.515$	&	$10.42$	\\
$0.05$	&	$6.269$	&	$78.36$	&	$6.341$	&	$67.11$	\\
$0.10$	&	$12.54$	&	$313.4$	&	$12.80$	&	$251.6$	
\end{tabular}
\caption{Coefficients of linear regression for $\tilde{\sigma}(\omega)$ in the interval $\sfrac{\omega}{\mathcal{K}v} \in [10^{-3},10^{-2}]$. }
\label{tab:Linear-regression}
\end{table}

As we see there is excellent agreement between the value of $\tilde{\sigma_0}$ at weak coupling and also a reasonable agreement for the value of $\tilde{\sigma_1}$ with those of the perturbative analysis of Ref. \cite{mishchenko2008minimal}. Therefore we have been able capture the logarithmic running of the coupling constant expected from the RG analysis at weak coupling.

\end{document}